\documentclass[preprint,12pt]{elsarticle}

\usepackage[table,xcdraw]{xcolor}
\usepackage{lineno} 
\usepackage{multirow}%
\usepackage{amsmath,amssymb,amsfonts,amsthm}%
\usepackage{mathrsfs}%
\usepackage[title]{appendix}%
\usepackage{textcomp}%
\usepackage{manyfoot}%
\usepackage{booktabs}%
\usepackage[ruled,vlined]{algorithm2e}
\usepackage{listings}%
\usepackage[utf8]{inputenc}

\usepackage{makecell}
\usepackage{parallel}
\usepackage{natbib}
\usepackage{mathtools}
\usepackage{fullpage}
\usepackage{lastpage}
\usepackage{enumitem}
\usepackage{wrapfig}
\usepackage{multicol}
\usepackage{cancel}
\usepackage{setspace}
\usepackage{calc}
\usepackage{siunitx}
\usepackage{xltabular}
\usepackage{threeparttablex}  % tables with footnotes

\usepackage{empheq}
\usepackage{framed}
\usepackage[most]{tcolorbox}
\usepackage[bold]{hhtensor}
\usepackage[T1]{fontenc}
\usepackage{hyperref}
\hypersetup{colorlinks = true, urlcolor   = blue, citecolor  = blue,}
\usepackage[caption=false]{subfig}  % subcaptions for subfigures
\usepackage{floatrow}
\newcommand{\mat}[1]{\boldsymbol{#1}}
\newcommand{\vect}[1]{\boldsymbol{#1}}

\journal{arXiv}

\begin{document}

\begin{frontmatter}
\title{ELEQTRONeX: A GPU-Accelerated Exascale Framework for Non-Equilibrium Quantum Transport in Nanomaterials}

\author[label1]{Saurabh S. Sawant\corref{cor1}}
\ead{saurabhsawant@lbl.gov}
\author[label2]{Fran\c{c}ois L\'eonard}
\ead{fleonar@sandia.gov}
\author[label1]{Zhi Yao}
\ead{jackie_zhiyao@lbl.gov}
\author[label1]{Andrew Nonaka}
\ead{ajnonaka@lbl.gov}

\cortext[cor1]{Corresponding author}

\affiliation[label1]{organization={The Center for Computational Sciences and Engineering, Lawrence Berkeley National Laboratory},%Department and Organization
            addressline={1 Cyclotron Road}, 
            city={Berkeley},
            postcode={94720}, 
            state={California}, 
            country={United States}}

\affiliation[label2]{organization={Sandia National Laboratories},%Department and Organization
            addressline={7011 East Ave}, 
            city={Livermore},
            postcode={94551}, 
            state={California}, 
            country={United States}}            

\begin{abstract}
Non-equilibrium electronic quantum transport is crucial for the operation of existing and envisioned electronic, optoelectronic, and spintronic devices. The ultimate goal of encompassing atomistic to mesoscopic length scales in the same nonequilibrium device simulation approach has traditionally been challenging due to the computational cost of high-fidelity coupled multiphysics and multiscale requirements. In this work, we present ELEQTRONeX (\textbf{ELE}ctrostatic \textbf{Q}uantum \textbf{TR}ansport modeling \textbf{O}f \textbf{N}anomaterials at \textbf{eX}ascale), a massively-parallel GPU-accelerated framework for self-consistently solving the nonequilibrium Green's function formalism and electrostatics in complex device geometries. By customizing algorithms for GPU multithreading, we achieve orders of magnitude improvement in computational time, and excellent scaling on up to 512 GPUs and billions of spatial grid cells. We validate our code by computing band structures, current-voltage characteristics, conductance, and drain-induced barrier lowering for various 3D configurations of carbon nanotube field-effect transistors.  We also demonstrate that ELEQTRONeX is suitable for complex device/material geometries where periodic approaches are not feasible, such as modeling of arrays of misaligned carbon nanotubes requiring fully 3D simulations.    
\end{abstract}

%\includegraphics[width=1\textwidth]{figures/GraphicalAbstract.png}

%\item ELEQTRONeX: an open-source self-consistent quantum transport framework for exascale systems
%\item MPI/GPU optimization in self-consistent NEGF method improves parallel time complexity
%\item ELEQTRONeX models 3D non-periodic carbon nanotube field effect transistors (CNTFETs)
%\item CNTFETs show robustness against misalignment and pitch variations in performance metrics
%\item Validation covers I-V characteristics, conductance, band-structures, drain-induced barrier lowering

%% Keywords

\end{frontmatter}
%\linenumbers

%\tableofcontents

% introduction
\clearpage
\section{Introduction}
\label{sec:introduction}

Non-equilibrium electronic transport determines the properties of many modern electronic devices.
As device dimensions are reduced and new nanomaterials introduced, novel quantum phenomena are being explored for improved performance.
In parallel, the inherent three-dimensional (3D) nature of nano- and meso-scale devices, including their constituent nanomaterials, requires a device modeling approach that can not only capture quantum phenomena but do so in a complex 3D geometry.
In particular, such devices may involve multiple densely packed materials/channels, necessitating modeling the cross-talk caused by neighboring channels to ensure accurate performance evaluation.
Thus, there is a need for an efficient approach capable of simulating multiple-channel devices on the order of hundreds of nanometer in length, while incorporating all of the atoms of the constituent materials, and capturing refined electrostatics over three orders of magnitude in size scale.

The Non-Equilibrium Green's Function (NEGF) formalism has emerged as a promising approach for device simulations.
It has been applied to the study of devices based on carbon nanotubes\citep{leonard_ballisticCNTFET_2006,leonard_crosstalk_2006}, graphene \citep{BANADAKI201580}, two-dimensional materials \citep{Ganapathi_2011}, nanowires \citep{Chowdhury_2023}, silicon nanosheets \citep{Park_2019}, and single molecules \citep{cohen_2020}. In addition it can describe time-dependent transport \citep{leonard_terahertz_2009} as well as response under external stimuli \citep{Leonard_graphene_photo_2017, Stewart_CNT_photo_2004}.
However, despite its versatility, NEGF simulations incur significant computational costs, primarily due to the need for self-consistently solving NEGF equations with electrostatics in complex 3D geometries.
While traditionally implemented through parallel deployment on CPUs, these simulations may still require hours to converge for a single device operating point \citep{Park_2019}.
Furthermore, this computational demand is further amplified when modeling multiple channels, as it necessitates simultaneous self-consistent computations of charge density across all channels.
Although numerous parallel NEGF implementations exist for CPUs, their GPU counterparts are notably scarce, with none currently capable of modeling multiple channels.

Sophisticated NEGF implementations include Transiesta~\citep{papior2017improvements,papior2016manipulating}, QuantumATK~\citep{smidstrup2019quantumatk}, NEMO5 \cite{klimeck2010atomistic, steiger2011nemo5}, the latter one is extended to GPUs using libraries like MAGMA and cuSPARSE~\cite{sahasrabudhe2014accelerating}, while other solvers use message passing interface (MPI) for parallelizations.
All of these implementations use independent computational units for parallel distribution such as energy points used for contour integration, discrete wavevectors, and voltage bias points.
However, this strategy suffers from scalability and memory limitations since each rank must compute the entire matrices and require storage of the entire non-distributed matrices on the order of tens of gigabytes in the processor memory.
These requirements can be alleviated by at most an order of magnitude using OpenMP threads per MPI rank~\citep{papior2017improvements, sahasrabudhe2014accelerating}. 
However, for a more substantial improvement and extension to modeling of multiple channels, a low-level optimization approach is needed that allows for multiple MPI ranks and GPUs to take part in computation of these matrices, say, at each energy point.

Initial attempts at low-level optimization for NEGF were made in Ref.~\cite{Jeong_2021} with a GPU-based implementation of the recursive Green's function (RGF) algorithm. However, this strategy is limited to decomposing the computations into two MPI ranks, each with its own GPU.
Others~\cite{libnegf} are leading efforts to port the libNEGF solver to GPUs using OpenACC. Currently, the implementation supports parallelization across energy and wavevector points, while ongoing investigations focus on domain decomposition.
These efforts have not yet addressed a parallelization strategy for self-consistently coupling with electrostatics, let alone extension to multiple channels.

We introduce ELEQTRONeX (\textbf{ELE}ctrostatic \textbf{Q}uantum \textbf{TR}ansport modeling \textbf{O}f \textbf{N}anomaterials at \textbf{eX}ascale), an open-source self-consistent NEGF implementation built on the DOE Exascale Computing Project AMReX library \cite{zhang19_AMReX,zhang21_AMReX}, emphasizing low-level MPI/GPU parallelization strategies across key components: electrostatics, NEGF, and the self-consistency algorithm.
The electrostatics module utilizes AMReX’s GPU-accelerated multigrid capabilities and can handle intricate shapes of terminal leads represented as embedded boundaries.
For NEGF parallelization, large matrices such as Green’s and spectral functions are decomposed on distributed across MPI ranks, with GPU acceleration applied to kernel computations.
We employ Broyden’s modified second method for self-consistency and present an MPI/GPU parallelization approach suitable for modeling multiple channel materials, simultaneously. 
Overall we demonstrate excellent scaling up to 512 GPUs.

In this work, we demonstrate the capability of ELEQTRONeX by modeling nanodevices with arrays of up to 20 carbon nanotubes of 10 and 100 nm channel lengths with varied average spacing. 
Our focus is on investigating the effect of angular misalignment and non-uniform spacing. 
These non-idealities closely reflect experimental device configurations~\cite{Arnold,Peng} and result in  non-periodic configurations, requiring fully 3D simulations.
By rigorously comparing several key device performance metrics, such as ON and OFF currents and subthreshold swing, against perfectly parallel configurations, we observe for the first time that carbon nanotube field effect transistors (CNTFETs) are robust against these non-idealities.
Additionally, we highlight that the modular design of the code allows for straightforward extension to model a wide range of materials beyond nanotubes.

The rest of this paper is organized as follows: \S \ref{sec:approach} describes our computational approach, highlighting MPI/GPU optimization strategies for each sub-module including electrostatics (\S \ref{subsec:electrostatics}), NEGF (\S \ref{subsec:negf}), and self-consistency (\S \ref{subsec:selfconsistency}). \S \ref{sec:validation} provides a rigorous validation of ELEQTRONeX for gate-all-around and planar CNTFET configurations. \S \ref{sec:scaling} demonstrates the parallel performance of the code for self-consistent simulations of gate-all-around CNTFET configurations using up to 512 GPUs and 400 nm channel length.
\S \ref{sec:scaling_results} discusses the current-voltage characteristics from the nonequilibrium simulations run for the parallel performance cases and highlights the challenges in simulating $O(100$~nm) nanotubes, self-consistently. 
In \S \ref{sec:nonpllPlanar}, we study the effect of modeling multiple CNTs in a fully 3D planar CNTFET configuration, focusing on the impact of CNT misalignment compared to perfectly parallel alignment, simulated with and without periodic boundary conditions.

%\clearpage
\section{Computational Approach}
\label{sec:approach}
\begin{figure}[tb]
    \centering
        \sidesubfloat[]{\label{f:GR_inverse}{\includegraphics[width=0.33\textwidth]{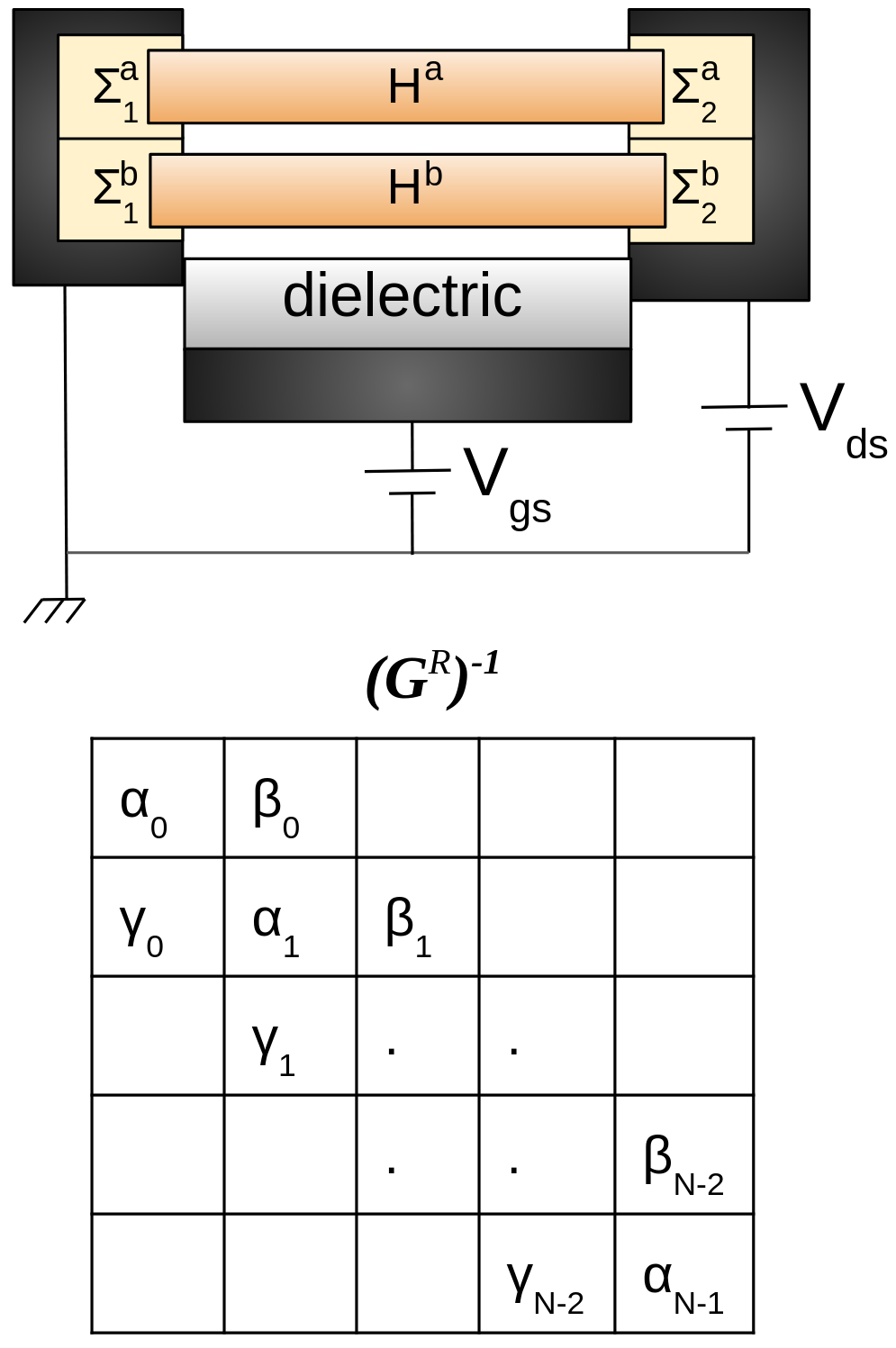}}}\hfill     
        \sidesubfloat[]{\label{f:Outline}{\includegraphics[width=0.45\textwidth]{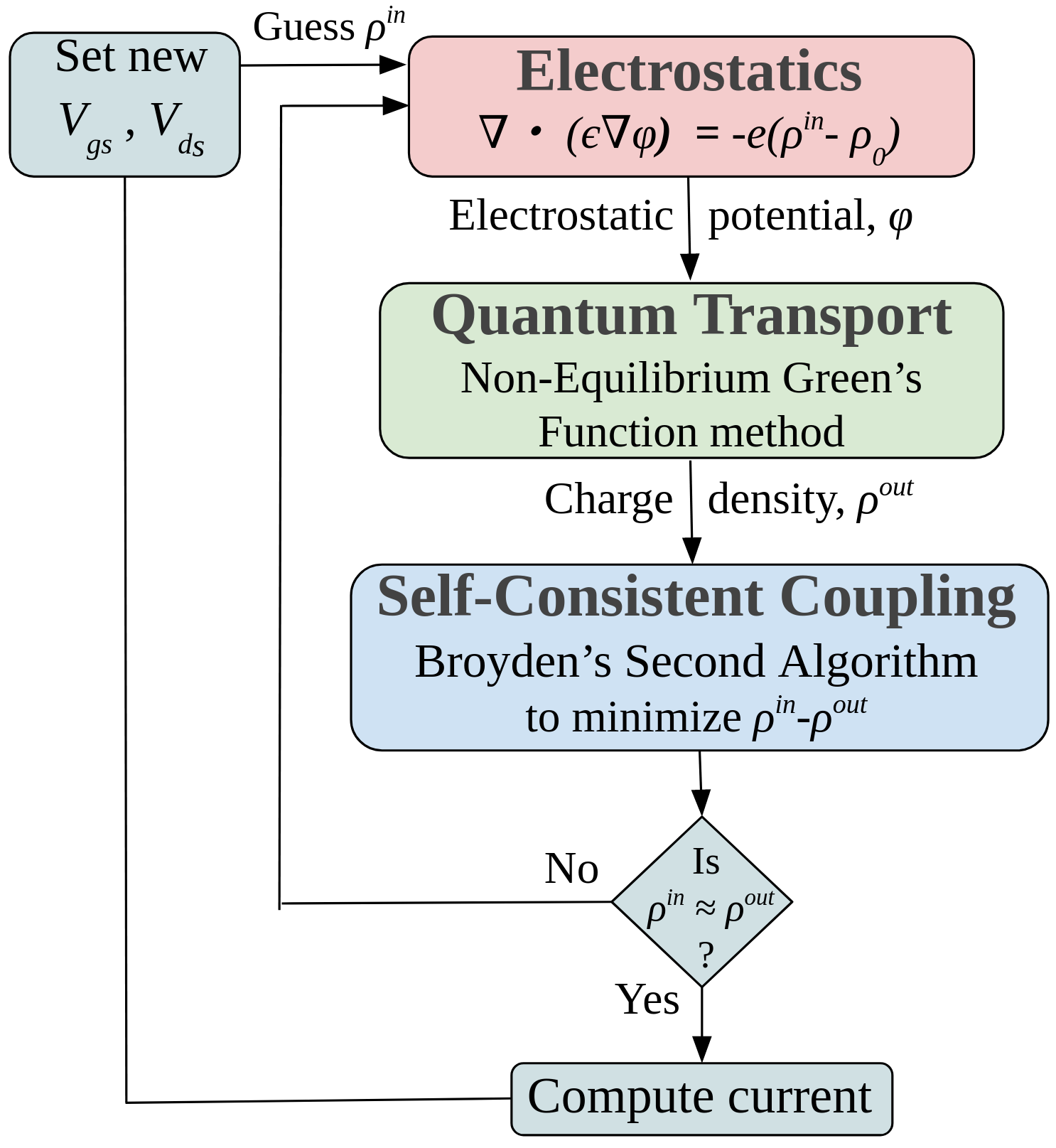}}}
        \caption{
        (\textit{a}) System setup for the NEGF method with multiple material channels, denoted by superscripts $a$ and $b$. The structure of the main matrix for computations is also shown. 
        (\textit{b}) Workflow for computing self-consistent quantum transport through nanomaterials.}
    \label{f:Sketch}
\end{figure}
In the application of the NEGF method for modeling quantum transport through nanodevices, the system is decomposed into material channels connected to semi-infinite external leads such as source and drain contacts, as described in Figure~\ref{f:GR_inverse}. Here, Hamiltonian $\mat{H}$ describes the electronic structure of each material channel, self-energies $\mat{\Sigma}_l^R$ for each lead $l$, represent its coupling to external leads, and $V_{gs}$ and $V_{ds}$ represent the gate-source and drain-source voltages, respectively.

To model the complete system, the ELEQTRONeX framework comprises three major components, as shown in Figure~\ref{f:Outline}: the electrostatic module, the quantum transport module, and the self-consistent coupling between the two modules.
The electrostatic module, described in \S\ref{subsec:electrostatics}, computes the electrostatic potential due to induced charges in the nanomaterial, as well as by device terminals such as source, drain, and gate, which can be modeled as embedded boundaries with intricate shapes.
Using the electrostatic potential on each site as input, the quantum transport module uses the NEGF method to compute retarded Green's function $\mat{G}^R(\vect{r},E)$ for each material channel, as described in \S\ref{subsec:negf}. This function can then be used to compute induced charges at the $N$ total sites in the material, also referred to as the system size. In this work, we model multiple carbon nanotubes using the mode-space approximation~\cite{guo2004toward}, where for each nanotube, $N$ corresponds to the total number of carbon rings along the length of the nanotube.
The induced charge computed by the NEGF method, in turn, affects the electrostatic potential, and therefore, the two must be solved iteratively until self-consistency is achieved.
This is achieved using Broyden's modified second algorithm, as described in \S\ref{subsec:selfconsistency}. 

In what follows, we provide a description of the major distinctive features of these modules, focusing on MPI and GPU parallelization. 
Each component description also includes a discussion of the algorithmic complexity and amenability to parallelization, with performance tests detailed in \S\ref{sec:scaling}.  
In our scaling tests we increase the nanotube length, making the number of carbon rings $N$ an appropriate parameterization of system size, as the number of grid cells and total atoms are proportional to $N$.  We describe our algorithmic components in terms of serial time complexity, as well as our parallel implementation time complexity assuming the number of computational resources increases in proportion to $N$.

All simulations in this work were performed using AMReX~\cite{amrex} (git hash $\le$ ae3af4339) and ELEQTRONeX~\cite{ELEQTRONeX} (git hash $\le$ 850e464).

\subsection{Electrostastics, Potential Interpolation, and Charge Deposition}
~\label{subsec:electrostatics}

The underlying spatial discretization is a finite-volume mesh with cuboid grid cells that store the electrostatic potential and permittivity.  Coupling with NEGF involves the use of static Lagrangian particles to model atoms/sites, which interact with the underlying spatial mesh through interpolation and deposition kernels.

The electrostatic module solves Poisson's equation,
\begin{equation}
    \nabla \cdot \left[ \epsilon (\vec{r}) \nabla \phi (\vec{r}) \right] = - e[\rho (\vec{r})-\rho_0 (\vec{r})]
\label{eq:Poisson}
\end{equation}
where $\epsilon (\vec{r})$ is the permittivity, $\phi (\vec{r})$ is the spatially varying electrostatic potential, $e$ is the electronic charge, $\rho (\vec{r})$ is the charge density, and $\rho_0 (\vec{r})$ is the background charge density.

To solve equation (\ref{eq:Poisson}) robustly, the electrostatic module uses the geometric multigrid-based linear solvers implemented in AMReX~\citep{zhang21_AMReX}.
The algorithm employs a series of V- or W-cycles, which are iterative procedures involving relaxation techniques, grid transfer operations, and coarse grid computations~\citep{amrex_userguide}.
Specifically, at each iteration, the algorithm transfers residual information between grid levels, applies relaxation methods to reduce errors, and performs computations at coarser grid levels~\citep{almgren1998conservative}.
Notably, at the coarsest level of the grid hierarchy, the discretized equations are directly solved using a strategy termed as the `bottom solver'. 
While the default choice for this task is BiCGSTAB (Bi-Conjugate Gradient STABilized)~\citep{bicgstab_vanderVorst}, users have the flexibility to select from a range of methods or integrate with external libraries such as hypre (High-Performance Preconditioners) for enhanced robustness~\citep{hypre}.
To handle complex, 3D device configurations, the solver capabilities include support for diverse domain boundary conditions as well as embedded boundaries (EBs, also referred to as cut-cell representations) to represent electric potential boundary conditions on the surfaces of intricate device terminal shapes.
Specifying a spatially-varying function of voltages on domain or EBs allows the computation of complete current-voltage characteristics of a device when coupled with NEGF.
Additionally, the solver supports heterogeneous permittivity distributions and allows precise specification of atom locations within computational cells, facilitated by electrostatic potential interpolation and charge deposition schemes based on the cloud-in-cell algorithm~\citep{birdsall1969_CIC}.
In this algorithm, the potential at each atomic location is obtained by tri-linear interpolation of the potential at eight neighboring cell-centered grid points. Similarly, the algorithm employs a tri-linear deposition scheme to distribute charge density from an atom to the eight neighboring grid points. Since we use the mode-space approximation for modeling carbon nanotubes, we average the potential computed over all atoms around the ring before using it in the NEGF. Similarly, we distribute the charge computed at each ring equally over all atoms around the ring.
Furthermore, this framework offers flexible domain decomposition strategies for parallelization, allowing the computational domain to be decomposed irrespective of EBs, regions of varying permittivity, and atomic structure of a material.

The serial time complexity of the linear solvers, as well as the charge deposition and potential interpolation algorithms, are $O(N)$.
These algorithms are highly parallelizable and scaling is bounded by inter-processor communication.  In an ideal parallel implementation, the parallel time complexity is $O(1)$.

\subsection{Quantum transport using the NEGF method}~\label{subsec:negf}
\vspace{-2em}
\subsubsection{Overview of the NEGF method}\label{sssec:overview}
In this work we represent Hamiltonian $\mat{H}$ of size $(N \times N)$ using a tight-binding model, but the scheme is amenable to any representation. 
The Hamiltonian has block tri-diagonal form, with the diagonal entries of $\mat{H}$ corresponding to the electrostatic potential energy $-e\phi$ at each site within the material; in the present case, at each carbon ring. 

Central to computing the DC transport properties of the device is the calculation of the retarded Green's function, expressed as
\begin{equation}
\begin{split}
    \mat{G}^R(\vec{r},E)  &= \left[ (E+ i\eta) \mat{I} - \mat{H}(\vec{r}) - \sum_{l}^{L} {\mat{\Sigma}_l^R} \right]^{-1}\\
    \mat{\Sigma}_l^R(E) &= \mat{\tau}_l^{\dagger} \mat{g}_l^R \mat{\tau}_l\\
\end{split}
\label{eq:GR}
\end{equation}
where $E$ denotes the electron energy, $\eta$ is an infinitesimal constant, $\mat{I}$ is the identity matrix, $L$ denotes the number of leads in the system, and $\mat{\tau}$ and $\mat{g}^R$ are matrices for the device-lead coupling and the surface Green's function, respectively. 
From equation~(\ref{eq:GR}) and Figure~\ref{f:GR_inverse}, we note that $\alpha_j = (E+i\eta) + H_{(j,j)} - \sum_l^L \Sigma_{l,(j,j)}$ for $j=0$ to $N-1$, and $\beta_j = H_{(j,j+1)}$, $\gamma_j = H_{(j+1,j)}$ for $j=0$ to $N-2$. 
For a system with 2 semi-infinite leads, only $\Sigma_{1,(0,0)}$ and $\Sigma_{2,(N-1,N-1)}$ are non-zero.

Using the retarded Green's function, we can compute charge density matrix $\mat{\rho}$ as
\begin{equation}
\begin{split}
    \mat{\rho}(\vec{r}) &=  -\frac{g_{s}g_{b}}{2\pi i} \int_{-\infty}^{\infty} \mat{G}^<(\vec{r}, E) dE \\
\end{split}
\label{eq:RhoGeneral}
\end{equation}
where $g_{s}=2$ and $g_{b}$ are spin and band degeneracies, $\mat{G}^< = i \sum_l^L \mat{A}_l F_l$ represents the lesser Green's function, $\mat{A}_l= \mat{G}^{R}\mat{\Gamma}_l \mat{G}^{R\dagger}$ denotes the lead-specific spectral function matrix, and $\mat{\Gamma}_l = i (\mat{\Sigma}_l^R - \mat{\Sigma}_l^{R\dagger})$ stands for the broadening matrix.
$F_l \equiv F_l(E-\mu_l, k_b T_l)$ represents the lead-specific Fermi function, depending on the electrochemical potential $\mu_l$ and temperature $T_l$, where $k_b$ is the Boltzmann constant.
Using the above identities and following the approach described in Refs.~\cite{taylor2001ab, brandbyge2002density, papior2017improvements}, We calculate $\mat{\rho}$ in two parts as
\begin{equation}
\begin{split}
    \mat{\rho}(\vec{r}) &=  \frac{1}{\pi} \Im \left[ \int_{-\infty}^{E_{min}}  \mat{G}^R(\vec{r},E) F_{min} \; dE \right]  -\frac{1}{2\pi} \int_{E_{min}}^{E_{max}} \sum_l \mat{A}_l(\vec{r},E) F_l \; dE \\
\end{split}
\label{eq:Rho}
\end{equation}
where $F_{min}$ is the lowest Fermi level, $E_{min}=\mu_{min} - f_1k_bT_{min}$, $E_{max} = \mu_{max} + f_2 k_bT_{max}$, ($\mu_{min}$, $T_{min}$) and ($\mu_{max}$, $T_{max}$) are minimum and maximum electrochemical potentials and temperatures, respectively. The factors $f_1$ and $f_2$ are problem-dependent and typically set to $14$ to capture the tail of the Fermi levels in the integration.
We efficiently calculate the first part of the integral using contour integration with the residue theorem, and use Gauss-Legendre mapping for accuracy with fewer integration points~\cite{papior2017improvements}. 
At equilibrium, i.e. when all leads have equal $\mu$ and $T$, the charge density can be calculated solely using the residue theorem by setting $E_l=E_u=\mu + fk_bT$.
Similarly, the neutral charge density matrix $\mat{\rho}_0(\vec{r})$  is computed with $E_l=E_u=0$.
The diagonal entries of the density matrix, $\mat{\rho}_{mm}$ and $\mat{\rho}_{0,mm}$, represent the charge at each material site, which are spread on the computational grid using the cloud-in-cell algorithm before recomputing potential using equation~(\ref{eq:Poisson}).

Equations~(\ref{eq:Poisson}), (\ref{eq:GR}), and (\ref{eq:Rho}) are evaluated until self-consistency is achieved, as described in \S\ref{subsec:selfconsistency}. Subsequently, essential device properties such as current and transmission can be determined as
\begin{equation}
\begin{split}
    I_l &= \frac{g_{s} e}{h} \int_{E_{min}}^{E_{max}} \text{Tr}\left[F_l\mat{\Gamma}_l \mat{A} + i \mat{\Gamma}_l \mat{G}^{<}  \right] dE  \\
    T_{pq}(E) &= \text{Tr}\left[\mat{\Gamma}_{p} \mat{G}^{R} \mat{\Gamma}_{q} \mat{G}^{R\dagger}\right] \\
\end{split}
\label{eq:deviceCharacteristics}
\end{equation}
where $I_l$ is the steady-state current through lead $l$, $h$ is Plank's constant, $\mat{A}=\sum_l^L \mat{A}_l$ is the total spectral function, and $T_{pq}$ is the transmission from lead $p$ to $q$. 
At zero source-drain bias, we may compute conductance as
\begin{equation}
\begin{split}
    G(E) &= G_0 \int_{-\infty}^{\infty} T(E) \left(-\frac{\partial F}{\partial E}\right) dE \\
\end{split}
\label{eq:zeroBiasG}
\end{equation}
where $T$ is the transmission at equilibrium and $G_0 = g_{s}e^2/h$ is the quantum of conductance.

%\begin{equation}
%\begin{split}
%    I_p &= \frac{2 e}{h} \sum_{m} \delta_m \int_{-\infty}^{\infty} \mat{\Gamma}_p^{m} \left(\mat{A}^m F_p + i \mat{G}^{<,m} \right) dE \\
%    T_{pq}(E) &= \sum_m \delta_m T_{pq}^{m} \\
%    T_{pq}^{m} &= Tr\left[\mat{\Gamma}_{p}^{m} \mat{G}^{R,m} \mat{\Gamma}_{q}^{m} \mat{G}^{R\dagger, m}\right] \\
%    G &= \frac{2e^2}{h} \sum_m \delta_m \int_{-\infty}^{\infty} T^m \left(-\frac{\partial F}{\partial E}\right) dE \\
%\end{split}
%\label{eq:deviceCharacteristics}
%\end{equation}
%where superscript $m$ stands for number of modes in the mode-space approximation, $I_p$ is the current through lead $p$, $\delta_m$ is the degeneracy for mode $m$, $T_{pq}^m$ is the transmission from lead $p$ to $q$ for mode $m$, 

\subsubsection{Flexible data structure to accommodate different materials}\label{sssec:dataStructure}
The composition of the matrices described above depends on the materials under consideration, the device structure, and the chosen representation for the Hamiltonian. 
Each block or element of these matrices can be a single number, a submatrix, or an array storing entries of a diagonal submatrix. 
For example, when modeling a CNT under the mode-space approximation with a single subband, each block is represented as a single number. However, when considering multiple subbands, it is represented as an array, with each entry corresponding to a subband.

To ensure flexibility in data structure and facilitate future extensions to other materials, we employed various C++ object-oriented programming techniques, including templates, virtual functions, and operator overloading. The general procedure of the NEGF method is implemented as a templated base class, with specialized classes derived from it for specific materials and their unique data structures. The general code in the base class, such as the one used to compute equation~(\ref{eq:GR}), works regardless of the underlying data structure of the materials-specific matrices. This is achieved by overloading mathematical operators, such as matrix multiplication, for the specific data structures of a material. Furthermore, during the execution of specific steps in the NEGF algorithm, the code searches for material-specific specializations of that step. If none are found, it defaults to the general procedure. For instance, the surface Green's function $g^R$ can be computed using a decimation technique~\cite{sancho_1985, wang2007nonequilibrium}, but may be overridden, for example when an analytical expression is available (e.g. for CNTs~\cite{guo2004toward}). This approach facilitates easy extension to new materials, as only the aspects specific to the new material need to be coded.

\subsubsection{Overview of parallelization strategy}\label{sssec:pllStrategy}
In the NEGF approach, the two core computations are the matrix inversion to compute the retarded Green's function $\mat{G}^R$ and the calculation of the spectral function $\mat{A}$. 
These computations are performed multiple times, for each integration point on all contour paths used to compute density matrix~(see equation~(\ref{eq:Rho})). 
Typically this means $\mathcal{O}(100-10k)$ of these computations per iteration of the self-consistency algorithm, depending on the problem.
Existing parallelization approaches map each energy point to a MPI process, leading to an efficient parallel scenario with no communication overhead for these tasks. 
However, this approach has limitations. 
Firstly, it restricts parallelism to the number of energy points, limiting scalability. 
Secondly, it imposes substantial memory demands, as each MPI process must store entire matrices for $\mat{G}^R$ and $\mat{A}$, alongside additional memory for other quantities.
This constraint can significantly restrict the utilization of available MPI processes, even on supercomputers.
For instance, on the Perlmutter supercomputer \cite{Perlmutter}, CPU-only nodes feature 64 cores with 512 GB DDR4 DRAM memory, allowing for the storage of only two complex floating-point matrices of size (16k × 16k).
For complex 3D configurations, we require a scalable solver capable of managing matrices substantially larger in size—by one or two orders of magnitude—while also allocating sufficient memory for computational grids related to electrostatics on each node. Additionally, any future extension of the code to model time-dependent equations will further increase the requirements for the number of matrices stored~\cite{leonard_terahertz_2009}.

To address these challenges and enable efficient computations on MPI and GPUs, we adopt a novel parallelization strategy. 
In our approach, each MPI process is assigned the responsibility of computing a nearly equal predetermined number of columns of the Green's or spectral function matrix. 
This distribution ensures that all MPI processes contribute to computations at each energy point, while significantly reducing memory requirements per process.
Consequently, each MPI process only needs to store Hamiltonian entries and compute charges for the assigned range of columns.
Moreover, this strategy facilitates a high degree of parallelization for large systems, when the number of columns ($N$) in the Hamiltonian matrix far exceeds the number of energy points.

In extending this strategy to GPUs, each MPI process is bound to a GPU device, with GPU threads invoked to handle the assigned columns, independently. The GPU-accelerated portion of the code is designed with portability in mind, ensuring compatibility across different GPU architectures and vendors. This approach maximizes computational efficiency while accommodating diverse hardware configurations. 

\subsubsection{Parallel computations of Green's and spectral functions}
\label{sssec:GandAcomput}

\begin{figure}[H]
    \centering
        \sidesubfloat[]{\label{f:TBD_a}{\includegraphics[width=0.19\textwidth]{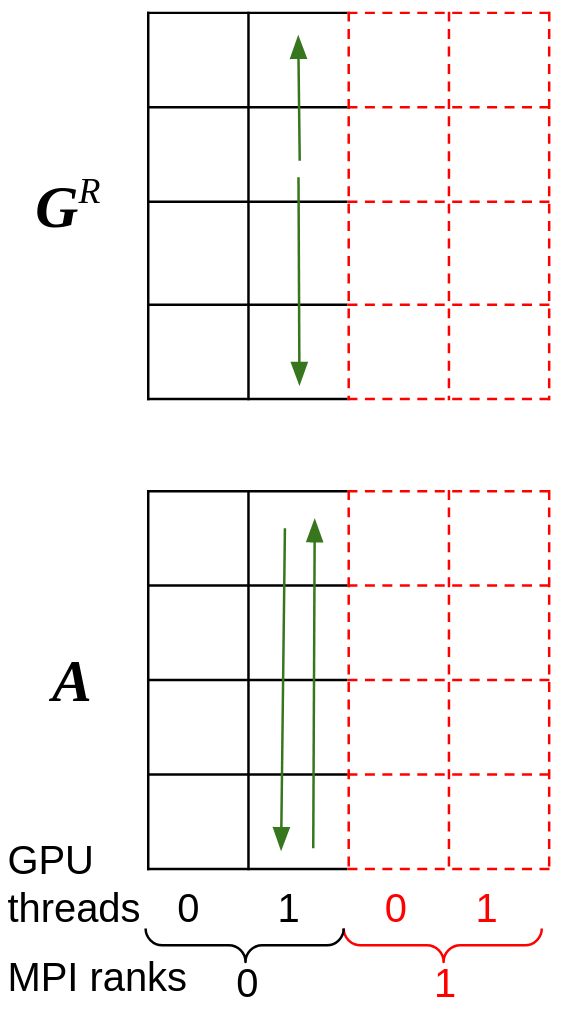}}}\hfill
        \sidesubfloat[]{\label{f:TBD_b}{\includegraphics[width=0.3\textwidth]{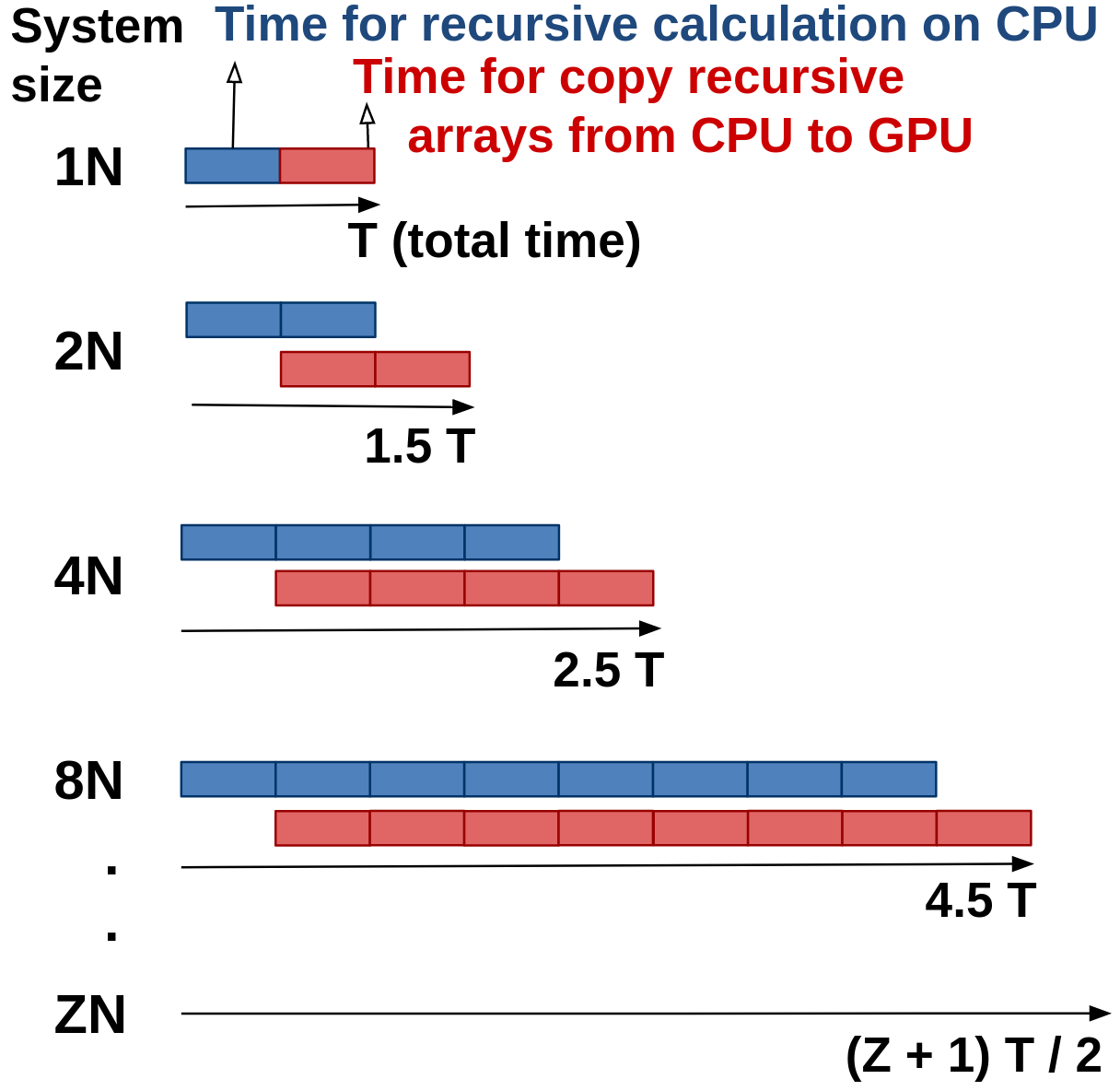}}}\hfill
        \sidesubfloat[]{\label{f:WVTF}{\includegraphics[width=0.36\textwidth]{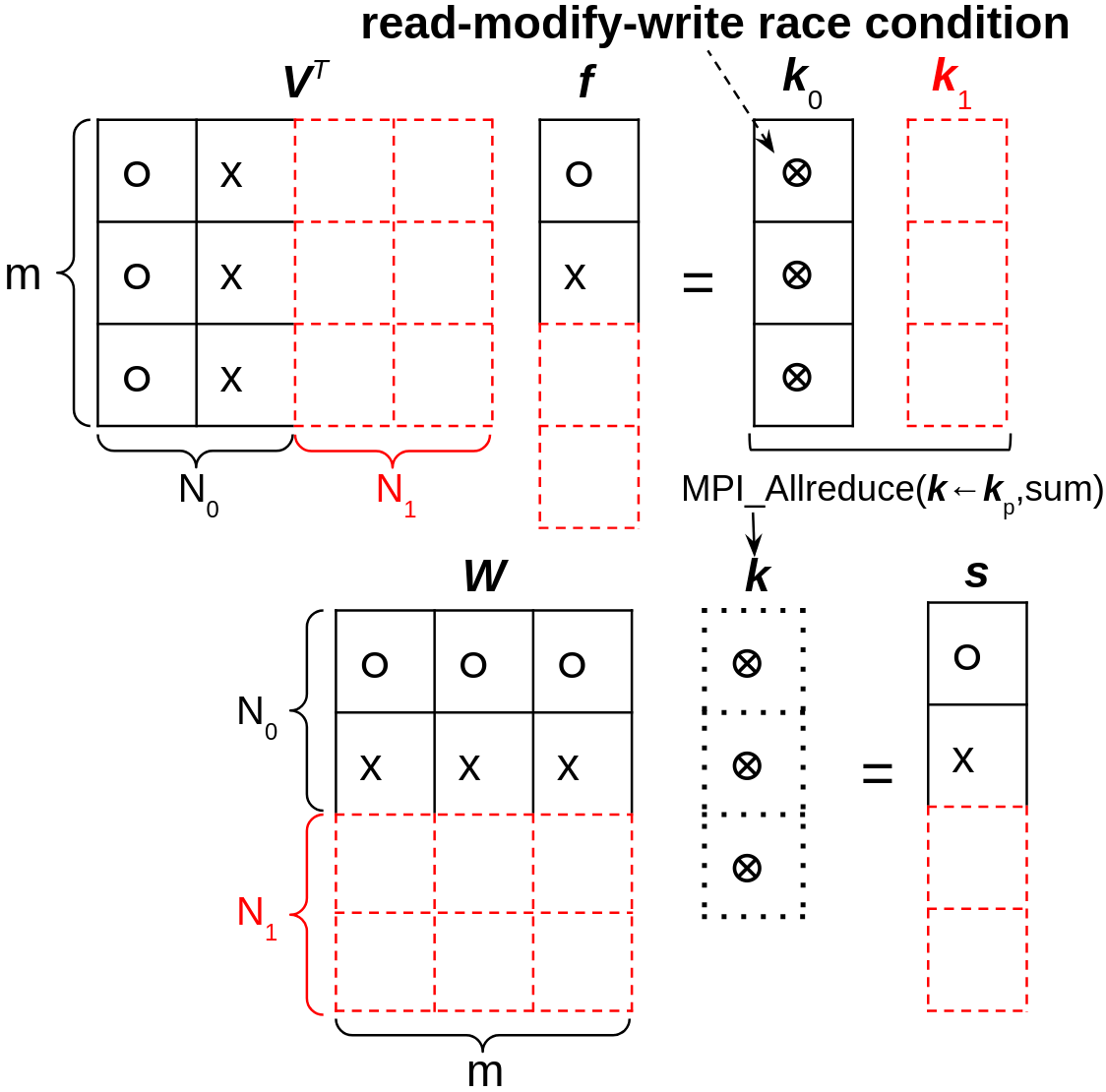}}}
        \caption{(\textit{a}) Illustration of the MPI/GPU parallelized part of the algorithm for computation of $\mat{G}^R$ and $\mat{A}$. (\textit{b}) Optimization of recursive part of the algorithm involving overlap of CPU computations and CPU-to-GPU asynchronous copy. (\textit{c}) Illustration of MPI/GPU parallelization strategy for steps 14 to 16 in Algorithm~\ref{Alg:Broyden} to compute $\vect{s}_p = \mat{W}_p \mat{V}_p^T \vect{f}_p$, assuming iterator $m$=3, total MPI ranks 2, and $N_0=N_1=2$. Solid and dashed edges of matrices and vectors distinguish elements stored and processed by MPI ranks 0 and 1, respectively, while the elements read or written by GPU threads $0$, $1$, or both $0$ and $1$, of MPI rank 0, are represented by a circle (o), cross (x), or by a circle-cross overlap, respectively.}
    \label{f:TBD}
\end{figure}

To compute $\mat{G}^R$ and $\mat{A}$, we employ a block tri-diagonal matrix inversion algorithm~\citep{godfrin1991method, hod2006first}, typically implemented as a serial algorithm in existing NEGF numerical methods~\citep{papior2017improvements}.
Here, we adopt this algorithm for hybrid MPI/GPU architectures with one GPU per MPI rank, such that individual columns of $\mat{G}^R$ and $\mat{A}$ are computed independently by parallelizing across all invoked MPI ranks and GPU threads, as illustrated in Algorithm~\ref{Alg:BTD_parallel} and Figure~\ref{f:TBD_a}.
For the computation of $\mat{G}^R$, each GPU-thread invoked on each MPI-rank begins by calculating the diagonal element of a column and subsequently computing the off-diagonal elements of that column using a recursive approach.
The spectral function $\mat{A}$ is computed by aggregating the lead-specific spectral functions $\mat{A}_l$, for each lead $l$, where each column of $\mat{A}_l$ is computed recursively, similarly to the Green's function.

\begin{algorithm}[H]
    \KwIn{$\vect{\alpha}, \vect{x}, \vect{\tilde{x}}, \vect{y}, \vect{\tilde{y}}$ as computed from Algorithm~\ref{Alg:BTD_recursive}.}
    \KwOut{$\mat{G}^R, \mat{A}$.}
    \texttt{Define:} Let $N_p$ be the portion of system size evaluated by MPI rank $p$.\\
    \texttt{Set:} $\mat{G}^R, \mat{A}$ to zero.
    
   \ForEach{\textnormal{\texttt{MPI rank $p$}}}{
   \ForEach{\textnormal{\texttt{GPU thread $t$ | $t \in [0, N_p)$}}}
   {
       \nl $T \leftarrow t + p N_p$\tcp*{global index, assuming $N_p$ is the same for all $p$}
       \nl $\mat{G}_{(T,T)}^R \leftarrow [\vect{\alpha}_{(T)} - \vect{x}_{(T)} - \vect{y}_{(T)}]^{-1}$\tcp*{diagonal element}
       \nl $\mat{G}_{(n+1,T)}^R \leftarrow -\vect{\tilde{x}}_{(n)} \mat{G}_{(n,T)}^R \quad \text{for } n \ge T$\tcp*{elements below diagonal}
       \nl $\mat{G}_{(n-1,T)}^R \leftarrow -\vect{\tilde{y}}_{(n)} \mat{G}_{(n,T)}^R \quad \text{for } n \le T$\tcp*{elements above diagonal}
    \ForEach{\textnormal{\texttt{lead $l$ | $l \in [1, L]$}}}
     {
       \nl \texttt{set: } $k \leftarrow \text{block index for lead } l$\;
       \nl $\mat{A}_{l,(n,T)} \leftarrow \mat{G}_{(n,k)}^R \Gamma_{l,(k,k)} \mat{G}_{(T,k)}^{R\dagger} \quad \text{for } n=k$\;
       \nl $\mat{A}_{l,(n+1,T)} \leftarrow -\vect{\tilde{x}}_{(n)} \mat{A}_{l(n,T)} \quad \text{for } n \ge T$\;
       \nl $\mat{A}_{l,(n-1,T)} \leftarrow -\vect{\tilde{y}}_{(n)} \mat{A}_{l,(n,T)} \quad \text{for } n \le T$\;
       \nl $\mat{A} \leftarrow \mat{A} + \mat{A}_{l}$\tcp*{aggregate $\mat{A}_l$ into $\mat{A}$.}
       }
   } }
  \caption{MPI/GPU parallelization of $\mat{G}^R$ and $\mat{A}$ computations.}
  \label{Alg:BTD_parallel}
\end{algorithm}

To initiate the recursion for $\mat{A}_l$, each GPU thread computes  $\mat{A}_{l,(n,T)}$, setting $n=k$, subsequently, computing the remaining entries in a given column using a suitable recursion method, as shown in Algorithm~\ref{Alg:BTD_parallel}. Here, $T$ represents the global index depending on MPI rank $p$ and GPU thread index $t$, and $k$ corresponds to the block index for lead $l$. To begin the recursion, each GPU thread needs to store some lead-specific quantities and perform some additional computations.
For example, in a system with two semi-infinite leads having block indices $k=0$ and $N-1$, respectively, the second GPU thread ($t=1$) on MPI rank $p=0$ initially computes entries $\mat{A}_{1,(0,1)}= \mat{G}_{(0,0)}^R \mat{\Gamma}_{1,(0,0)} \mat{G}_{(1,0)}^{R\dagger}$ and $\mat{A}_{2,(N-1,1)} = \mat{G}_{(N-1,N-1)}^R \mat{\Gamma}_{2,(N-1,N-1)} \mat{G}_{(1,N-1)}^{R\dagger}$. 
To achieve this, the GPU thread needs lead specific quantities, $\Gamma_{1,(0,0)}$ and $\Gamma_{2,(N-1,N-1)}$, and entries $\mat{G}_{(0,0)}$, $\mat{G}_{(N-1,N-1)}$, which can in turn be used to compute, $\mat{G}_{(1,0)}$ and $\mat{G}_{(1,N-1)}$ using a suitable recursion.

In practice, for NEGF method with tight-binding approximation for Hamiltonian, we only need the diagonal entries of $\mat{G}^R$ and $\mat{A}$, and therefore, we offer a way to turn off storage of off-diagonal elements of these matrices. 
However, computing the diagonal elements of $\mat{A}_l$ still requires elements $\mat{G}_{(T,k)}^R$, necessitating $O(N)$ recursive computations per thread.

%\begin{figure}[H]
%    \centering
%        \sidesubfloat[]{\label{f:TBD_a}{\includegraphics[width=0.45\textwidth]{figures/approach/GR_and_A.png}}}\hfill
%        \sidesubfloat[]{\label{f:TBD_b}{\includegraphics[width=0.45\textwidth]{figures/approach/async_optimization.png}}}\,
%        \caption{(\textit{a}) Illustration of the MPI/GPU parallelized part of the algorithm for computation of $\mat{G}^R$ and $\mat{A}$. (\textit{b}) Illustration of the recursive part of the algorithm, and optimization of the algorithm involving overlap of CPU computations and CPU-to-GPU asynchronous copy. }
%    \label{f:TBD}
%\end{figure}

From Figure~\ref{f:TBD_a}, we see that to perform aforementioned recursions in parallel, each MPI-rank and GPU-thread needs access to vectors $\vect{\tilde{x}}$ and $\vect{\tilde{y}}$.
These vectors can be computed using a recursion shown in Algorithm~\ref{Alg:BTD_recursive} using vectors $\vect{\alpha}$, $\vect{\beta}$, $\vect{\gamma}$, defined earlier in \S~\ref{sssec:overview} and Figure~\ref{f:GR_inverse}, representing the diagonal, upper and lower adjacent diagonals of the block-tridiagonal matrix $(G^{R})^{-1}$ that is being inverted.

\begin{algorithm}[H]
    \KwIn{Block vectors $\vect{\alpha}, \vect{\beta}, \vect{\gamma}$ as defined in Figure~\ref{f:GR_inverse}.}
    \KwOut{Block vectors $\vect{x}, \vect{\tilde{x}}, \vect{y}, \vect{\tilde{y}}$ of size $N$.}
   \texttt{Set:} $\vect{y}_{(0)} = \vect{x}_{(N-1)}=0$.\\
   \ForEach{\textnormal{\texttt{vector index \(n\) | \(n \in [0, N)\)}}}
   {
   $\vect{y}_{(n)} \leftarrow \vect{\beta}_{(n-1)} \vect{\tilde{y}}_{(n)}$\;
   $\vect{\tilde{y}}_{(n)} \leftarrow [\vect{\alpha}_{(n-1)} - \vect{y}_{(n-1)}]^{-1} \vect{\gamma}_{(n-1)}$\;
   $\vect{x}_{(n)} \leftarrow \vect{\gamma}_{(n)}  \vect{\tilde{x}}_{(n)}$\;
   $\vect{\tilde{x}}_{(n)} \leftarrow [\vect{\alpha}_{(n+1)} - \vect{x}_{(n+1)}]^{-1} \vect{\beta}_{(n)}$\;
   }
  \caption{Recursive computations of vectors needed as input for Algorithm~\ref{Alg:BTD_parallel}.}
  \label{Alg:BTD_recursive}
\end{algorithm}

The recursive computation of $\vect{\tilde{x}}$ and $\vect{\tilde{y}}$ poses a bottleneck for this algorithm, as explained shortly, limiting its time complexity to be proportional to $O(N)$, regardless of number of MPI ranks used.
This limitation can be partially mitigated by implementing long recursion in smaller segments and overlapping parts of recursion computations with CPU-to-GPU asynchronous copies, as illustrated in Figure~\ref{f:TBD_b}.
For instance, assuming a system size $N$ simulated with $P$ total MPI ranks, doubling the system size to $2N$ would typically result in double of the algorithm time, even if computational resources are doubled to $2P$ MPI ranks.
Yet, by overlapping the computation of the latter half of the array with the CPU-to-GPU copy of the former half, the algorithm time can be reduced to a factor of approximately $1.5$.
Scaling this approach to simulate system size $ZN$ leads to a time-increase factor of $(Z+1)/2$, instead of $Z$.
In practice, the exact value of this factor depends on the size of the recursive array segment chosen for overlap optimization.
In general, if the segment size is such that the time for computation equals the CPU-to-GPU copy time, we may approach the ideal limit.

The time complexity of the serial block-tridiagonal algorithm is $O(N^2)$, whereas the parallel time complexity can be expressed as,
\begin{equation}
T(N) = O\left(\frac{N}{s_1}\right) + O\left(\frac{N^2}{s_2 P}\right).
\end{equation}
Here, the first and second terms represent time complexities of the serial CPU-based recursion and GPU-parallelizable parts of the algorithm, respectively, $s_1 \in [1,2)$ represents the speedup obtained from the overlap optimization discussed above, $s_2$ represents the speedup of GPUs over CPUs in evaluating a column of length $N$ (typically, $s_2 \sim O(100)$), and $P$ indicates the total number of parallel processing units.
In our GPU implementation we invoke as many GPU threads as the number of columns per processor, thus $P$ is typically equal to $N$ and the resulting parallel time complexity is $T(N) = O(N/s_1) + O(N/s_2)$.
Since $s_1 < s_2$, we expect the algorithm to scale as $O(N/s_1)$, and $s_1$ to approach 2 in the best case scenario.
%On the other hand, without GPUs, the time complexity becomes $T(N) = O(N) + O(N^2/P)$, and since typically, $P<<N$, we expect the second term to dominate, highlighting the advantage of using GPUs.

In the future, we plan to pre-compute and store the vectors $\vect{\tilde{x}}$ and $\vect{\tilde{y}}$ for each energy point over which $\mat{G}^R$ and $\mat{A}$ are computed. 
This way the linear cost of recursion can be further amortized over the number of independent energy points, leading to potentially significant savings in time, albeit at the cost of additional memory requirement on the order of $O(NE_{\rm pts})$, where $E_{\rm pts}$ is the number of energy points.

\subsection{Parallel self-consistency module using Broyden's second algorithm}~\label{subsec:selfconsistency}
%\vspace{-2em}

The self-consistent calculation involves finding the nearest zero of the function $\vect{f} = \vect{\rho}^{in} - \vect{\rho}^{out}$ where the charge density reaches a self-consistent value. 
The superscripts `$in$' and `$out$' denote the input charge densities used for computing the electrostatic potential and those generated from the NEGF algorithm, respectively.
Note that for this algorithm we only need the diagonal entries of the density matrix.
To achieve faster convergence, we employ Broyden's modified second algorithm, which does not require the storage of the inverse Jacobian matrix, thus significantly reducing memory usage.
Since the algorithm details are provided in various references \citep{srivastava1984broyden, singh1986accelerating, ihnatsenka2007electron, areshkin2010electron}, we focus solely on presenting the MPI/GPU parallelization of this algorithm here.

The pseudo-code for the MPI/GPU parallelization of this algorithm is outlined in Algorithm~\ref{Alg:Broyden}. 
Adhering to a strategy consistent with the parallelization of the NEGF algorithm, each MPI rank is assigned to compute a designated subset of the charge density vector, denoted as $\vect{\rho}_p$, with a size of $N_p$, where the subscript $p$ denotes quantities specific to MPI rank $p$. 
Step 1 of the algorithm involves using $\vect{\rho}_p^{in}$ and $\vect{\rho}_p^{out}$ as input to electrostatics and output from the NEGF algorithm for MPI rank $1$, respectively. 
Steps 2 to 4 entail calculating local vectors, which can be easily parallelized across GPU threads. 
Steps 5 and 6 involve reduction operations, where all threads attempt to modify a single variable, resulting in a read-modify-write race condition~\citep{wen2022programming}.
To address this, GPU-atomic operations are utilized to ensure that the read-modify-write operation sequence on a memory location by one thread does not overlap with that by the other thread. 
Additionally, more advanced reduction algorithms such as warp-shuffle~\cite{nvidia_warpshuffle} may offer improved efficiency, although they were not implemented in this work.
Steps 7 and 8 involve reduction operations across all MPI ranks, which can be achieved using the \texttt{MPI\_Allreduce} primitive.

Steps 9 to 16 are intermediate steps to compute the vector $\vect{s}_p$, which is used to update the input charge density $\vect{\rho}_p^{in}$ for the next iteration.
Among these, steps 12, 13, 17, and 18 can be easily parallelized across GPU threads without requiring any synchronization or communication across GPU threads or MPI ranks.
The most computationally intensive steps are steps 9 to 11 and steps 14 to 16, which are very similar in nature.
Here, we provide elaboration on steps 14 to 16, which entail computation of $\vect{s}_p = \mat{W}_p \mat{V}_p^T \vect{f}_p$, using the illustration provided in Figure~\ref{f:WVTF}.

We begin with computing the matrix-vector product, $\vect{k}_p = \mat{V}_p^T \vect{f}_p$, where each thread is assigned to process one column of $\mat{V}_p^T$.
However, this step introduces a read-modify-write race condition, when multiple threads (in this case, 2) attempt to modify the same elements of $\vect{k}_0$ concurrently, similar to steps 5 and 6.
To address this issue, we employ GPU-atomic operations to ensure proper synchronization.
The subsequent step involves aggregating all local copies of $\vect{k}_p$, into a global copy $\vect{k}$ on each MPI rank using the $\texttt{MPI\_Allreduce}$ primitive.
Finally, in the third step, we evaluate $\vect{s}_p = \mat{W}_p \vect{k}$, where each thread is responsible for computing one row of $\mat{W}_p$ multiplied by $\vect{k}$.
The computational time complexity of the serial algorithm is $O(m_{\rm avg} N)$, where $m_{\rm avg}$ represents the average number of iterations required for convergence for a system size of $N$.
In an ideal implementation the parallel time complexity is $O(m_{\rm avg})$, noting that as the system size increases $m_{\rm avg}$ increases sub-linearly in our demonstration cases.
%The computational complexity of the serial algorithm is $O(m_{\rm avg} N)$; as we scale the problem size in $N$ and increase the computational resources in proportion we would expect an increase in run-time proportional to the increase in $O(m_{\rm avg})$.

This algorithm can be easily extended to accommodate simulations involving multiple channel materials without requiring modifications to the core algorithm.
In this scenario, $N_p$ represents the sum of portions of all system sizes evaluated by MPI rank $p$, where $N_p$ is defined as $\sum_{c=1}^{c=C} N_p^c$, where $C$ represents number of material channels, and $N_p^c$ represents portion of the system size of material channel $c$ evaluated by MPI rank $p$.
For example, $\vect{\rho}_p^{in}$, represents a vector composed of all charge density vectors $\vect{\rho}_p^c$ for $c \in [1, C]$, denoted as $\vect{\rho}_p^{in} \equiv (\vect{\rho}_p^{in,1}, \vect{\rho}_p^{in,2}, ..., \vect{\rho}_p^{in,C})$.
This array can be formed using the $\texttt{MPI\_Allgatherv}$ primitive before proceeding to step $2$.
The subsequent steps ($2$ to $18$) are executed as usual.
Before repeating step $1$ for the next iteration, we decompose the vector $\vect{\rho}_p^{in}$ into channel-specific charge densities $\vect{\rho}_p^{in,c}$ for all $c \in [1, C]$.
This strategy ensures that during minimization of the function $\vect{f} = \vect{\rho}^{in} - \vect{\rho}^{out}$ we account for contribution from all material channels.

\begin{algorithm}[H]
\footnotesize % Reduce font size for the algorithm
\setstretch{0.9} % Adjust line spacing within the algorithm
\vspace{-0.1em}
    \KwIn{A guess for charge density $\vect{\rho}^{in}$, Broyden fraction $\zeta$, maximum number of iterations, $M$.}
    \KwOut{Self-consistent charge $\vect{\rho}^{out} \approx \vect{\rho}^{in}$.}
    \texttt{Define:} $\mat{V}_p^T$ and $\mat{W}_p$ as matrices of size $M \times N_{p}$ and $N_{p} \times M$, respectively, $\vec{k}_p$ as vector of size $M$, and all other quantities denoted by lowercase bold letters as vectors of size $N_p$, where $N_{p}$ is the portion of system  size evaluated by MPI rank $p$.\\
    \texttt{Set:} iterator, $m \leftarrow 0$, maximum relative norm, $\epsilon_{max} \leftarrow \infty$, and all matrices, vectors to $0$.

    \While{$\epsilon_{max} < 10^{-5}$ \textnormal{\texttt{and}} $m < M$ }{
    \ForEach{\textnormal{\texttt{MPI rank $p$}}}{
 \nl    $\vect{\rho}_p^{in} \rightarrow \text{Electrostatics} \rightarrow \text{NEGF} \rightarrow \vect{\rho}_p^{out}$\;
 %         \ForEach{\textnormal{\texttt{GPU thread $t \epsilon (0, N_p)$}}}
          \ForEach{\textnormal{\texttt{GPU thread \(t\) | \(t \in [0, N_p)\)}}}
          {
              \nl $\vect{f}_{p,(t)} \leftarrow \vect{\rho}_{p,(t)}^{in} - \vect{\rho}_{p,(t)}^{out}$\tcp*{compute minimization function} 
              \nl  $\vect{\Delta f}_{p,(t)} \leftarrow \vect{f}_{p,(t)} - \vect{f}_{p,(t)}^{\text{old}}$\tcp*{compute change in minimization function}
              \nl  $\vect{\epsilon}_{p,(t)} \leftarrow \left| \vect{f}_{p,(t)}/(\vect{\rho}_{p,(t)}^{in} + \vect{\rho}_{p,(t)}^{out}) \right|$\tcp*{compute relative norm}
              \nl $\epsilon_{max,p} \overset{\text{atomic}}{\leftarrow} \max(\vect{\epsilon}_{p,(t)})$\tcp*{assign maximum element of $\vect{\epsilon}_p$ to $\epsilon_{max,p}$}
              \nl     $d_p \overset{\text{atomic}}{\leftarrow} ||\vect{\Delta f}_{p,(t)}||^2$\tcp*{aggregate square of each element of $\vect{\Delta f}_p$ into $d_p$}
          }
 \nl \texttt{MPI\_Allreduce}($\epsilon_{max} \leftarrow \epsilon_{max,p}$, max)\tcp*{find maximum $\epsilon_{max,p}$ over all processes}
 \nl \texttt{MPI\_Allreduce}($d \leftarrow d_p$, sum)\tcp*{aggregate sum of $d_p$ over all processes}
    \If{m > 0} 
    {
        \If{m > 1} 
        {
            \ForEach{\textnormal{\texttt{GPU thread \(t\) | \(t \in [0, N_p)\)}}}
            {
                \ForEach{\textnormal{\texttt{index \(j\) | \(j \in [1, m-1]\)}}}
                {
                    \nl $\vect{k}_{p,(j)} \overset{\text{atomic}}{\leftarrow} \mat{V}_{p,(j,t)}^T \vect{\Delta f}_{p,(t)}$\tcp*{read-modify-write race condition}
                }
            }
            \nl \texttt{MPI\_Allreduce}($\vect{k} \leftarrow \vect{k}_p$, sum)\tcp*{aggregate sum of $\vect{k}_p$ over all processes}
            \ForEach{\textnormal{\texttt{GPU thread \(t\) | \(t \in [0, N_p)\)}}}
            {
                \ForEach{\textnormal{\texttt{index \(j\) | \(j \in [1, m-1]\)}}}
                {
                    \nl $\vect{s}_{p,(t)} \leftarrow \mat{W}_{p,(t,j)} \vect{k}_{(j)}$\tcp*{each thread reads $\vect{k}$}
                }
            }
        }
        \ForEach{\textnormal{\texttt{GPU thread \(t\) | \(t \in [0, N_p)\)}}}
        {
           \nl $\mat{V}_{p,(m,t)}^T \leftarrow d^{-1}\vect{\Delta f}_{p,(t)}$\tcp*{Evaluate $m^{th}$ row of $\mat{V}_p^T$}
           \nl $\mat{W}_{p,(t,m)} \leftarrow - \eta \vect{\Delta f}_{p,(t)} + \vect{\rho}_{p,(t)}^{in} - \vect{\rho}_{p,(t)}^{in,old} - \vect{s}_{p,(t)}$\tcp*{Evaluate $m^{th}$ column}
        }

            \ForEach{\textnormal{\texttt{GPU thread \(t\) | \(t \in [0, N_p)\)}}}
            {
                \ForEach{\textnormal{\texttt{index \(j\) | \(j \in [1, m]\)}}}
                {
                    \nl $\vect{k}_{p,(j)} \overset{\text{atomic}}{\leftarrow} \mat{V}_{p,(j,t)}^T \vect{f}_{p,(t)}$\tcp*{read-modify-write race condition}
                }
            }
 \nl    \texttt{MPI\_Allreduce}($\vect{k} \leftarrow \vect{k}_p$, sum)\tcp*{aggregate sum of $\vect{k}_p$ over all processes}
            \ForEach{\textnormal{\texttt{GPU thread \(t\) | \(t \in [0, N_p)\)}}}
            {
                \ForEach{\textnormal{\texttt{index \(j\) | \(j \in [1, m]\)}}}
                {
                    \nl $\vect{s}_{p,(t)} \leftarrow \mat{W}_{p,(t,j)} \vect{k}_{(j)}$\tcp*{each thread reads $\vect{k}$}
                }
            }

    }
       \ForEach{\textnormal{\texttt{GPU thread \(t\) | \(t \in [0, N_p)\)}}}
       {
           \nl   $\vect{\rho}_{p,(t)}^{in, old} \leftarrow \vect{\rho}_{p,(t)}^{in}$\tcp*{update $\vect{n}_{p,(t)}^{in, old}$ for next iteration}
        \nl   $\vect{\rho}_{p,(t)}^{in} \leftarrow \vect{\rho}_{p,(t)}^{in,old} - \zeta \vect{f}_{p,(t)} - \vect{s}_{p,(t)}$\tcp*{predict $\vect{n}_{p,(t)}^{in}$ for next iteration}
      }
    }
    \nl   $m \leftarrow m+1$\tcp*{increment iterator $m$}
  }
  \caption{MPI/GPU Parallelization of Broyden's modified second algorithm.}
  \label{Alg:Broyden}
\end{algorithm}

\section{Results}
%\clearpage

Here we provide validation of ELEQTRONeX for gate-all-around and planar CNTFET configurations (\S \ref{sec:validation}). Then we demonstrate the parallel performance of the code for self-consistent simulations of gate-all-around CNTFET configurations (\S \ref{sec:scaling}).
Next, we discuss the current-voltage characteristics from these nonequilibrium simulations and discuss challenges in modeling longer nanotubes (\S \ref{sec:scaling_results}).
Finally, we study the effect of modeling multiple CNTs, both aligned and non-aligned, in fully 3D planar CNTFET configurations compared to corresponding periodic configurations (\S \ref{sec:nonpllPlanar}).

\subsection{Validation Studies}
\label{sec:validation}

We validate ELEQTRONeX in obtaining DC transport properties by comparing our results with those reported in Refs.~\cite{leonard_ballisticCNTFET_2006} and \cite{leonard_crosstalk_2006} for gate-all-around and planar CNTFET configurations, respectively. 
Further details specific to modeling CNTFETs can be found in the references. In brief, we consider a (17,0) zigzag CNT of diameter 1.3 nm and nearest-neighbor tight-binding coupling of 2.5 eV, giving a bandgap of 0.55 eV. The Hamiltonian is written in terms of nearest-neighbor coupling between parallel carbon rings. The Fermi level of the source and drain contacts is 1 eV below the CNT midgap before self-consistency, as would be expected from Pd. The gate oxide has a dielectric constant of 3.9 to simulate SiO$_2$.

\begin{figure}[H]
    \centering
        \sidesubfloat[]{\label{f:L20nm_I_Vds}{\includegraphics[width=0.45\textwidth]{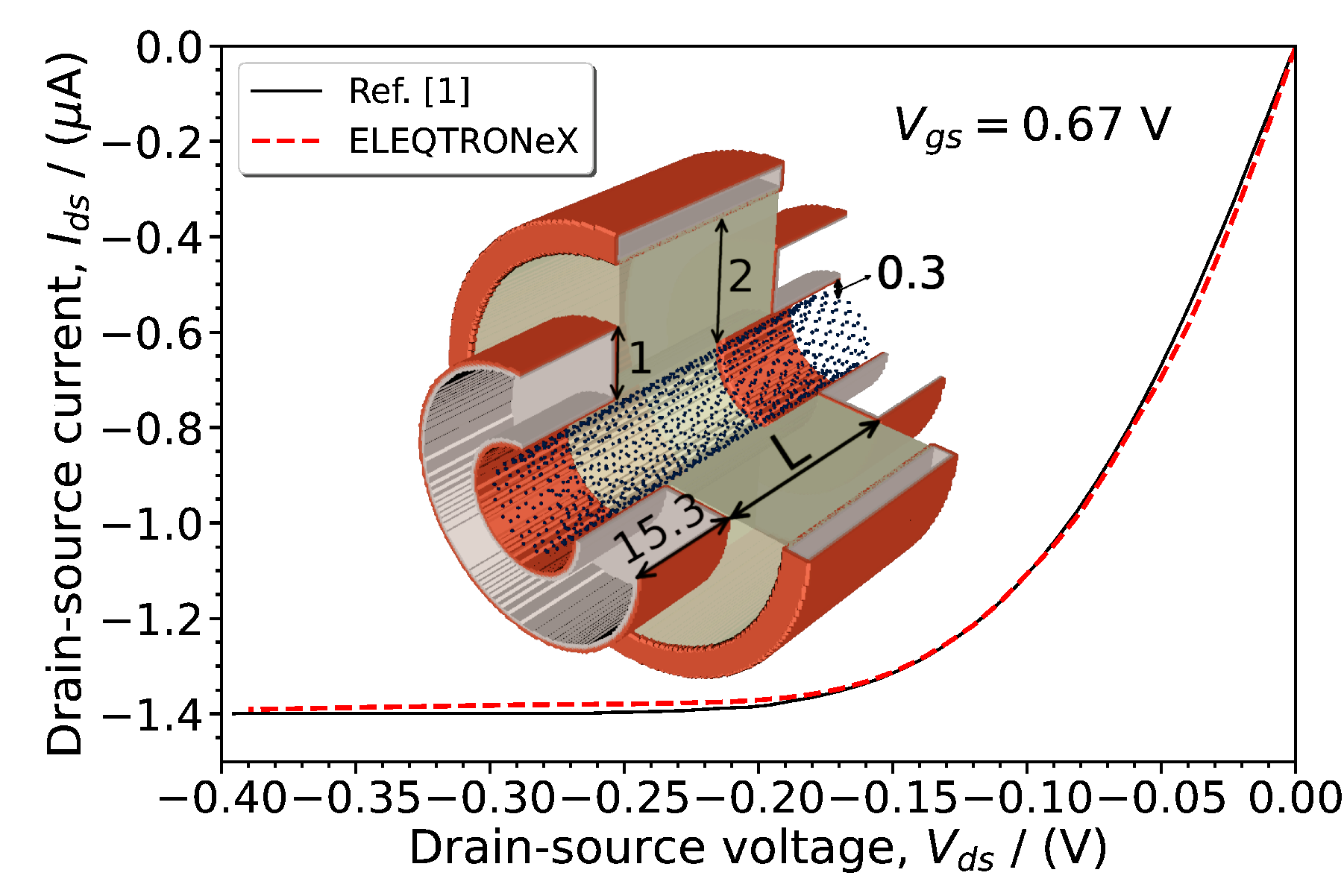}}}
        \hfill
        \sidesubfloat[]{\label{f:L20nm_I_Vgs}{\includegraphics[width=0.45\textwidth]{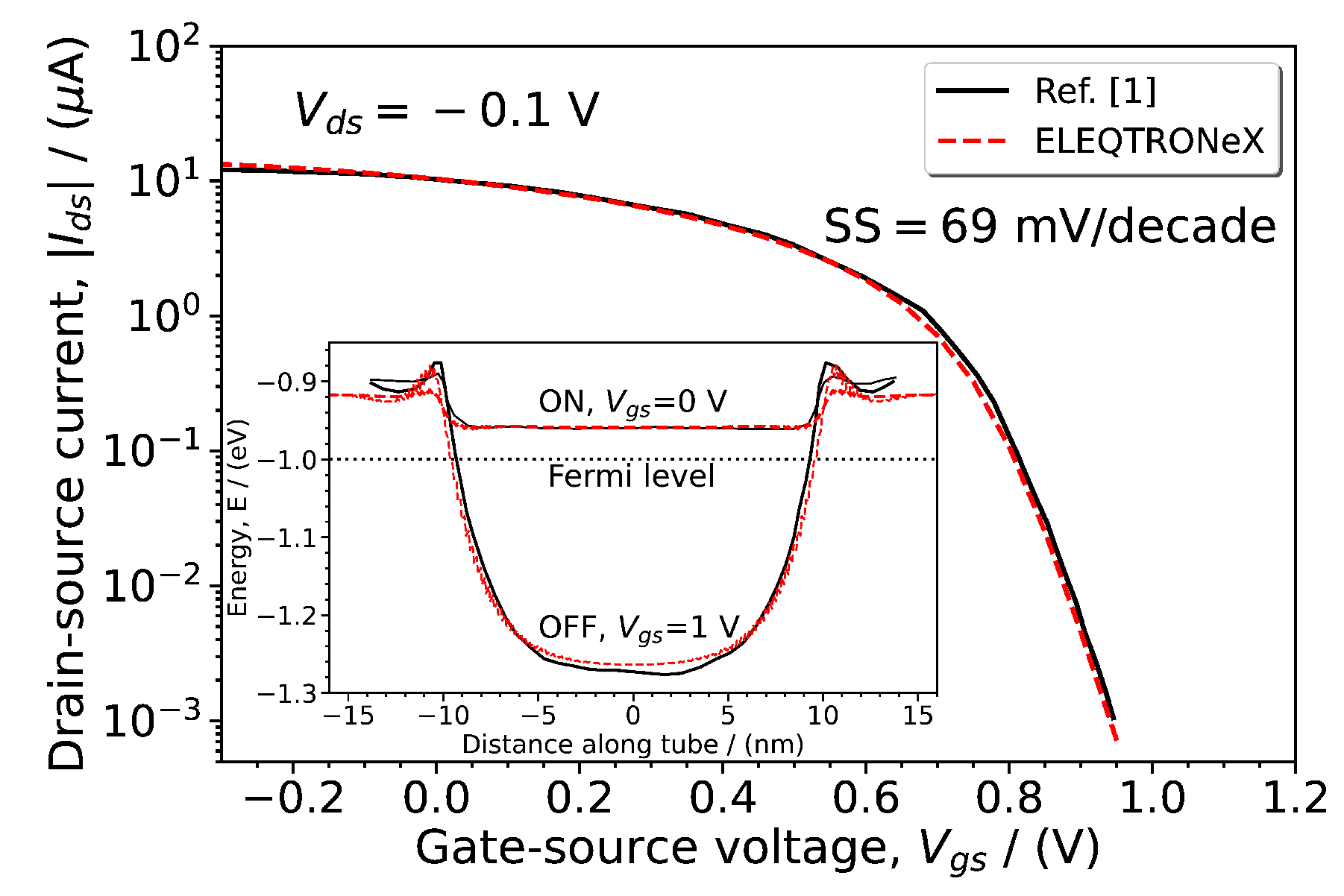}}}
        \,
        \sidesubfloat[]{\label{f:DIBL}{\includegraphics[width=0.45\textwidth]{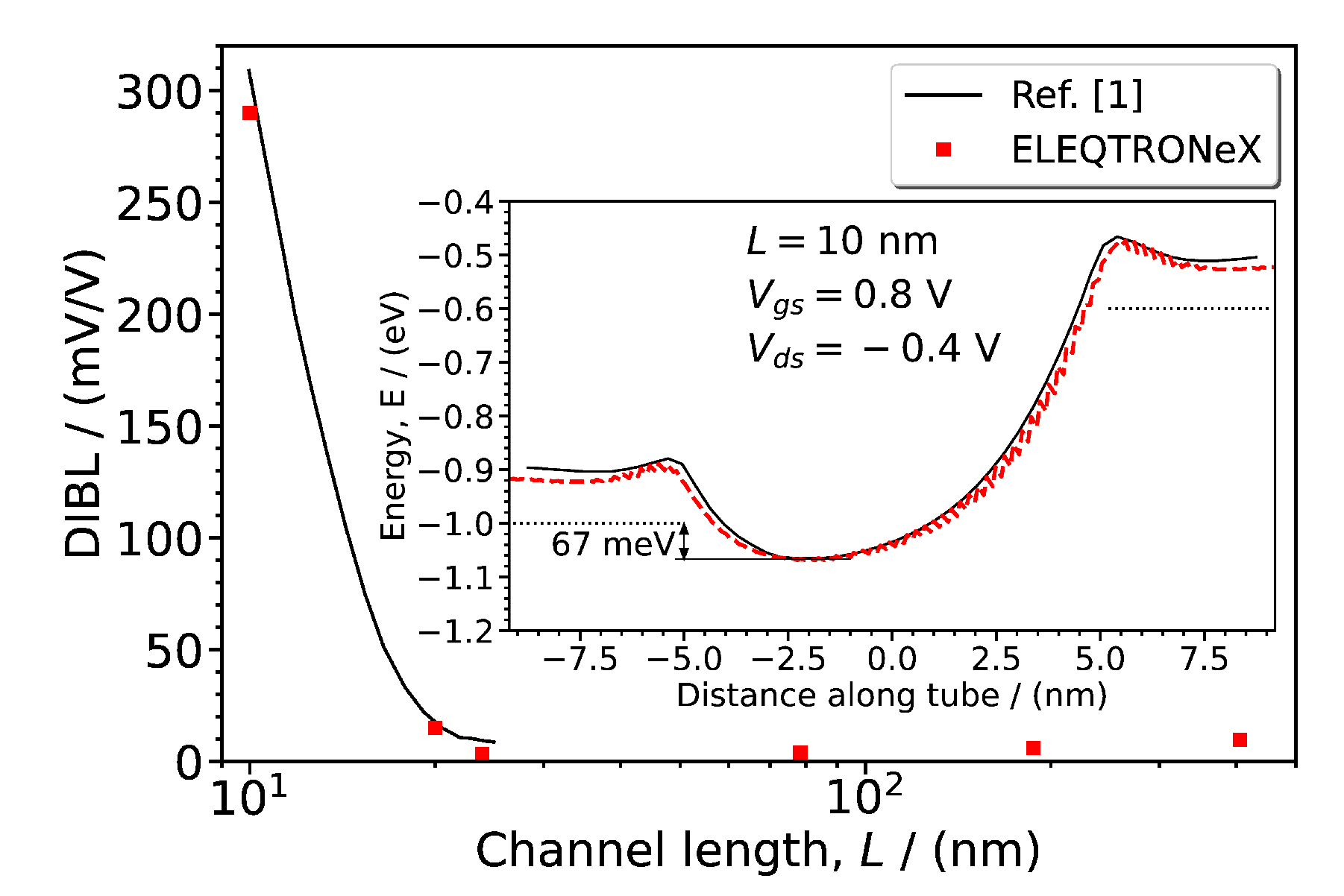}}}
        \hfill
        \sidesubfloat[]{\label{f:L10nm_periodic_G}{\includegraphics[width=0.44\textwidth]{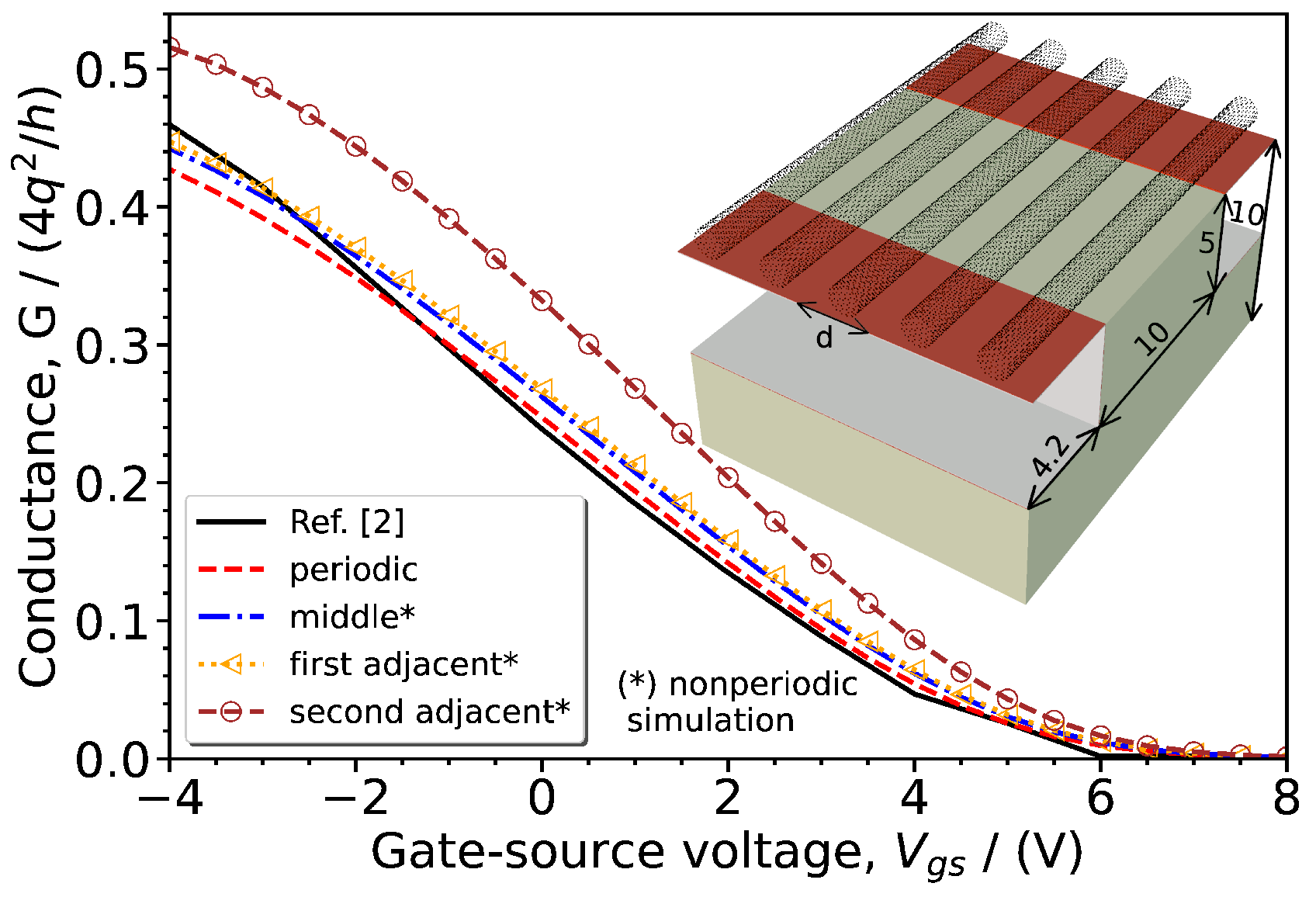}}}
        \,       
        \caption{
            Current-voltage characteristics of CNTFETs. (\textit{a})  Current versus source-drain voltage for a gate-all-around CNTFET with channel length of $L=20$ nm. (\textit{b}) Current versus gate-source voltage for the gate-all-around CNTFET. (\textit{c}) Drain-induced barrier lowering (DIBL) as a function of $L$. (\textit{d}) Conductance $G$ versus $V_{gs}$ for planar CNTFET configurations with $d=3.2$ nm. In all panels insets show valence band edge and schematics with dimensions in nanometers. Comparison is made with the data in Refs \cite{leonard_ballisticCNTFET_2006} and \cite{leonard_crosstalk_2006}.
        }
    \label{f:validation}
\end{figure}

Figures~\ref{f:L20nm_I_Vds}, \ref{f:L20nm_I_Vgs}, and \ref{f:DIBL} demonstrate good agreement for the drain-source current $I_{ds}$ versus drain-source voltage $V_{ds}$, $I_{ds}$ versus gate-source voltage $V_{gs}$, and drain-induced barrier lowering (DIBL) versus channel length $L$, respectively, for a gate-all-around CNTFET configuration.
The simulations agree despite methodological differences compared to Ref.~\cite{leonard_ballisticCNTFET_2006}, including differences in solving Poisson's equation (finite-difference, successive overrelaxation method versus finite-volume, multigrid), charge deposition to mesh (Gaussian spreading versus cloud-in-cell), calculation of surface Green's function (iterative layer doubling technique~\citep{sancho_1985} versus analytical~\citep{guo2004toward}), and the type of Broyden's algorithm employed (first versus parallelized modified second method).
Furthermore, in Figure~\ref{f:DIBL}, DIBL remains flat for channel lengths exceeding 20 nm, consistent with the semi-empirical fit reported by Ref.~\cite{leonard_ballisticCNTFET_2006}.
These additional runs are discussed in later in \S \ref{sec:scaling_results} on performance studies.

Next, we validate the code's capability to model multiple channels in a full 3D geometry by simulating an array of nanotubes and comparing it with the results of \cite{leonard_crosstalk_2006} for a planar CNTFET configuration, as shown in Figure~\ref{f:L10nm_periodic_G}. 
Initially, we model a single nanotube with periodic boundary conditions in the lateral direction (domain width $d=3.2$~nm) to emulate an infinite array of nanotubes. 
The obtained conductance ($G$) from this simulation compares well with the results of \cite{leonard_crosstalk_2006}. 
%However, at lower gate voltages, when the channel is open, transport behavior depends on how the NEGF boundary conditions are modeled, leading to slight discrepancies possibly due to differences in the calculation of the surface Green's function.
Subsequently, we simulate five nanotubes (domain width $5d$), once again imposing periodic boundary conditions, to ensure consistent behavior of conductance for each nanotube. 
This validates the capability of our model to accurately simulate multiple nanotubes.
Later in \S \ref{sec:nonpllPlanar}, we will discuss the results from modeling only five CNTs to illustrate how finite-size effects can impact device performance.

%\clearpage
\subsection{Parallel Performance}
\label{sec:scaling}

In this section, we conduct performance studies of ELEQTRONeX in parallel using up to 512 MPI ranks with one GPU per MPI rank, focusing on how the code runtime is impacted as computational resources increase in proportion to system size $N$.
We define computational load by the length of the nanotube simulated; increase in the load involves increasing the channel length as well as computational cells used for electrostatic calculation, proportionately.

We first simulate a gate-all-around CNTFET [see inset of Figure~\ref{f:L20nm_I_Vds}] with a carbon nanotube of 54.53~nm overall length ($L$=23.9 nm) using 64 GPUs, followed by simulations with 2, 4, and 8 times longer nanotubes, employing a proportional increase in the number of GPUs.
We analyze changes in 'time per iteration' across three key components of the code—electrostatics, NEGF, and self-consistency. 
These times and other parameters are summarized in Table~\ref{tab:weak-scaling}.
For this study, we used a number of integration points ($\sim$12k) that was determined based on the need to obtain numerically converged solution for the largest channel length, as explained in detail in \S\ref{sec:scaling_results}.
These simulations were carried out using GPU resources at the National Energy Research Scientific Computing Center's (NERSC's) Perlmutter supercomputer~\cite{Perlmutter}, where each node consists of four NVIDIA A100 GPUs.

\begin{table}[H]
\begin{threeparttable}
\caption{Parameters used and times obtained from the weak scaling study.}
\label{tab:weak-scaling}       % Give a unique labela
        \begin{tabular}{|l|c|c|c|c|}
\hline\noalign{\smallskip}
   $^*$Channel length of nanotube modeled / (nm)              & 23.9 & 78.4 & 187.4 & 405.6 \\
   Number of carbon rings (system size), $N$                                          & 512 & 1024 & 2048 & 4096 \\
   Number of MPI ranks and GPUs used                          & 64   & 128  & 256  & 512   \\
   Computational cells in the length-wise direction    & 1536 & 3072 & 6144 & 12288 \\
   Average number of Broyden's iterations for convergence, $m_{\rm avg}$        & 21   & 22   & 28  & 46 \\\hline
   Time for electrostatics / (s)                            & 1.45 & 1.54 & 1.59 & 1.63  \\
   Time for interpolation of potential from mesh to atoms / (ms) & 3.0  & 4.0  & 3.4  & 5.8  \\
   Time for charge deposition to mesh / (ms)                     & 0.63  & 0.67  & 0.79  & 1.20  \\
   Time for NEGF / (s)                                     & 2.04 & 3.12 & 5.28 & 8.76  \\
   Time for self-consistency / (ms)                              & 0.50  & 0.59  & 0.74  & 1.00  \\
\noalign{\smallskip}\hline\noalign{\smallskip}
\end{tabular}
    \begin{tablenotes}
    \item [$^{*}$] Contact length was kept constant to 15.336~nm, computational cells in the transverse direction were $192 \times 192$, and approximate number of integration points used were 12k.
    \end{tablenotes}
\end{threeparttable}
\end{table}

Table~\ref{tab:weak-scaling} illustrates that the time for electrostatics remains nearly constant indicating near-perfect parallelization of the electrostatic components.
The small 12\% increase in the largest case could be attributed to increase in the overall communication time.
The time for the NEGF module increases proportionally to $N/2$, consistent with parallel time complexity discussed in \S \ref{sssec:GandAcomput}, where a factor of $1/2$ is achieved due to overlapping a portion of the recursive array computations with simultaneous CPU-to-GPU asynchronous copying. 
In terms of absolute times, Table~\ref{tab:weak-scaling} reveals that the times per iteration for Broyden's algorithm, interpolating the electrostatic potential from mesh to atomic sites, and depositing charge from atomic locations to mesh, are orders of magnitude smaller compared to that for electrostatics and NEGF.
Among these, the self-consistency module with the Broyden's algorithm exhibits time increase roughly proportional to the average number of iterations required for convergence, $m_{\rm avg}$, as expected.

In the above calculation, we used only one doubly-degenerate transport mode. 
Therefore, next, we demonstrate the effect of increasing the number of modes on computational time using a simpler simulation involving a CNT embedded in an all-around circular lead, same as the CNT-lead portion of the gate-all-around configuration shown in Figure~\ref{f:L20nm_I_Vds}. 
We vary the potential on the lead metal from to 0 to 1~V, while at each condition, the Broyden iterations are carried out to achieve self-consistency.
Since this is an equilibrium Green's function calculation, we simulate a very long CNT (55.84 $\mu$m), and vary the number of modes from 1 to 8 (all doubly-degenerate). 
For modes greater than one, each block of the block-tridiagonal matrix is represented in the form of an array with each array element corresponding to a mode.
Table~\ref{tab:weak-scaling-equilibrium} shows that with increase in number of modes from 1 to 8, the time for NEGF increases by only 2.84 times.
The sub-linear increase in time may be caused by two major reasons that benefit calculations on both CPUs and GPUs: better cache and memory bandwidth utilization, since the data for each mode is stored contiguously in an array, and better core-pipeline latency hiding due to increase in the load per thread.

\begin{table}[H]
\begin{threeparttable}
\caption{Parameters and times for the CNT-lead equilibrium calculations with multiple transport modes.}
\label{tab:weak-scaling-equilibrium}       % Give a unique labela
        \begin{tabular}{|l|c|c|c|c|}
\hline\noalign{\smallskip}
   Number of carbon rings (system size), $N$             & \multicolumn{4}{c|}{524,288}\\
   Computational cells          & \multicolumn{4}{c|}{$72 \times 1,572,864 \times 72$}\\
   Contour integration points   & \multicolumn{4}{c|}{$90$}\\\hline
   Number of modes              & 1 & 2 & 4 & 8 \\
   Time for NEGF / (s) & 1.98 & 2.75 & 3.24 & 5.63  \\\hline
   Time for electrostatics / (s) & \multicolumn{4}{c|}{2.2 - 2.4}  \\
\noalign{\smallskip}\hline\noalign{\smallskip}
\end{tabular}
\end{threeparttable}
\end{table}

\subsection{Long-Channel Gate-All-Around CNTFET Configurations}~\label{sec:scaling_results}

Here, we discuss the results obtained from the nonequilibrium simulations run for the parallel performance cases with gate-all-around configurations, discussed in Table \ref{tab:weak-scaling}, as well as some considerations in choosing integration paths and points to obtain converged self-consistent solution for longer nanotubes.

\begin{figure}[H]
    \centering
        %\sidesubfloat[]{\label{f:scaling}{\includegraphics[width=0.46\textwidth]{figures/scaling/scaling.eps}}}\hfill
        \sidesubfloat[]{\label{f:WeakResults_I_Vds}{\includegraphics[width=0.45\textwidth]{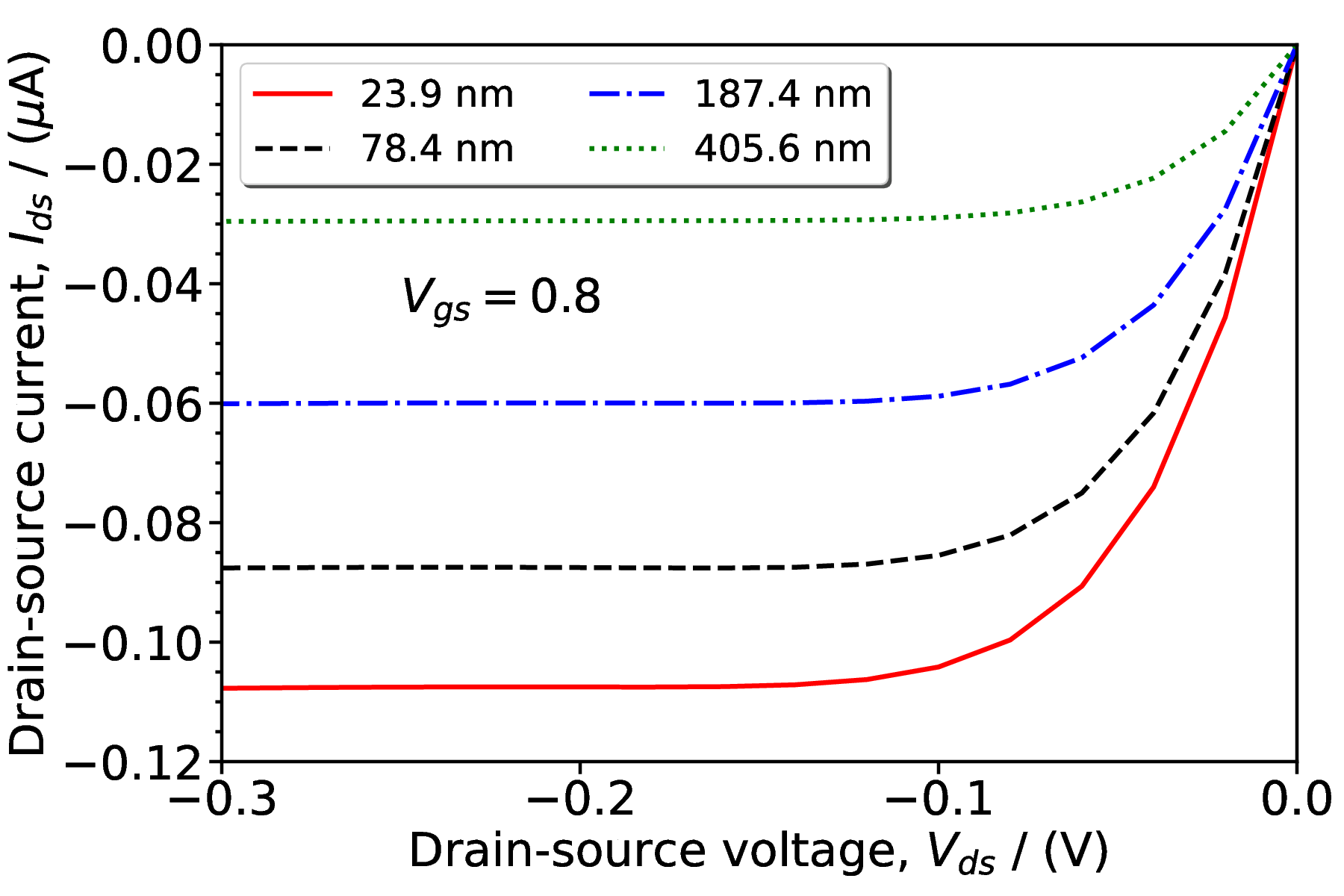}}}\hfill
        \sidesubfloat[]{\label{f:WeakResults_I_Vgs}{\includegraphics[width=0.45\textwidth]{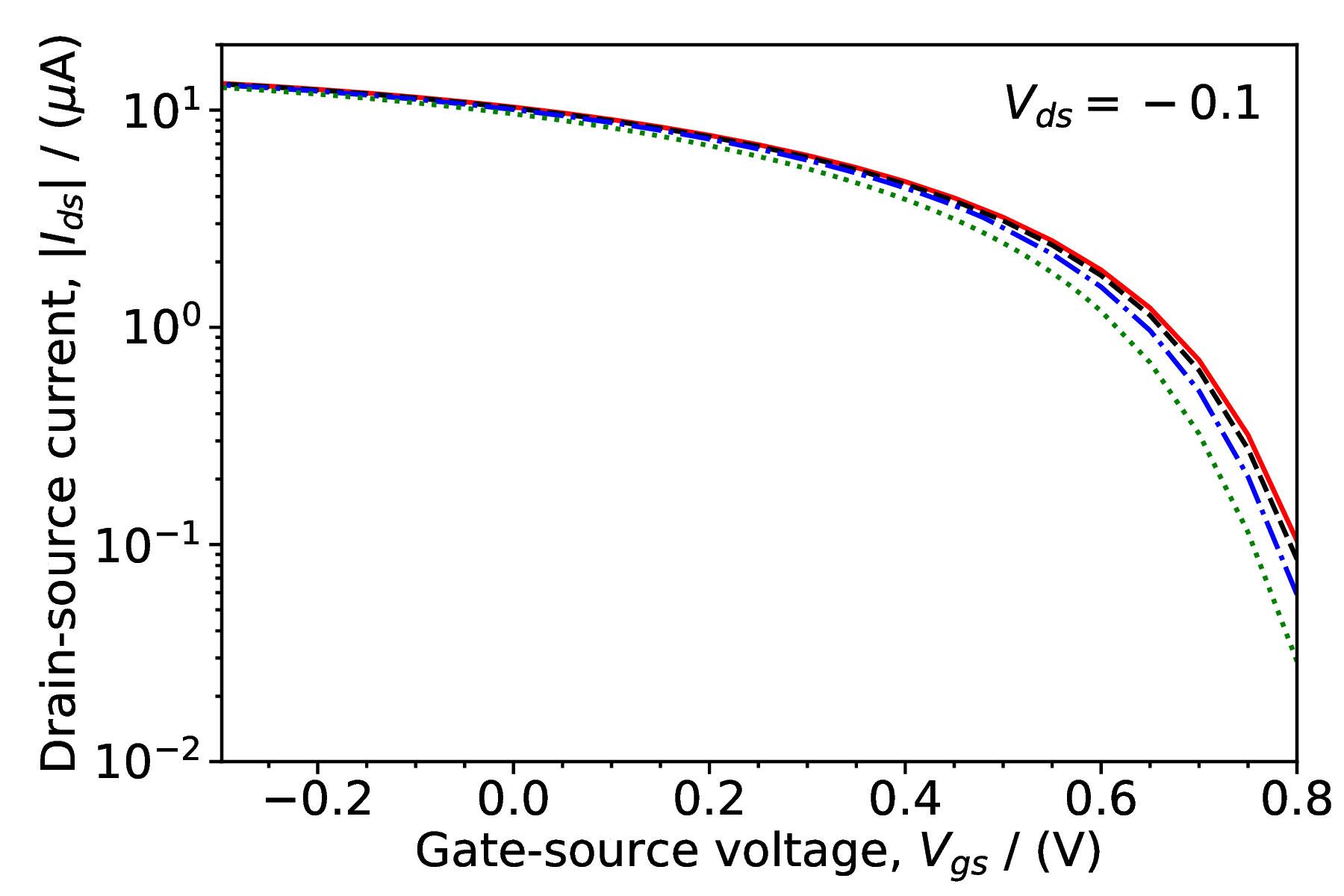}}}\,
        \sidesubfloat[]{\label{f:WeakResults_bandStructure}{\includegraphics[width=0.45\textwidth]{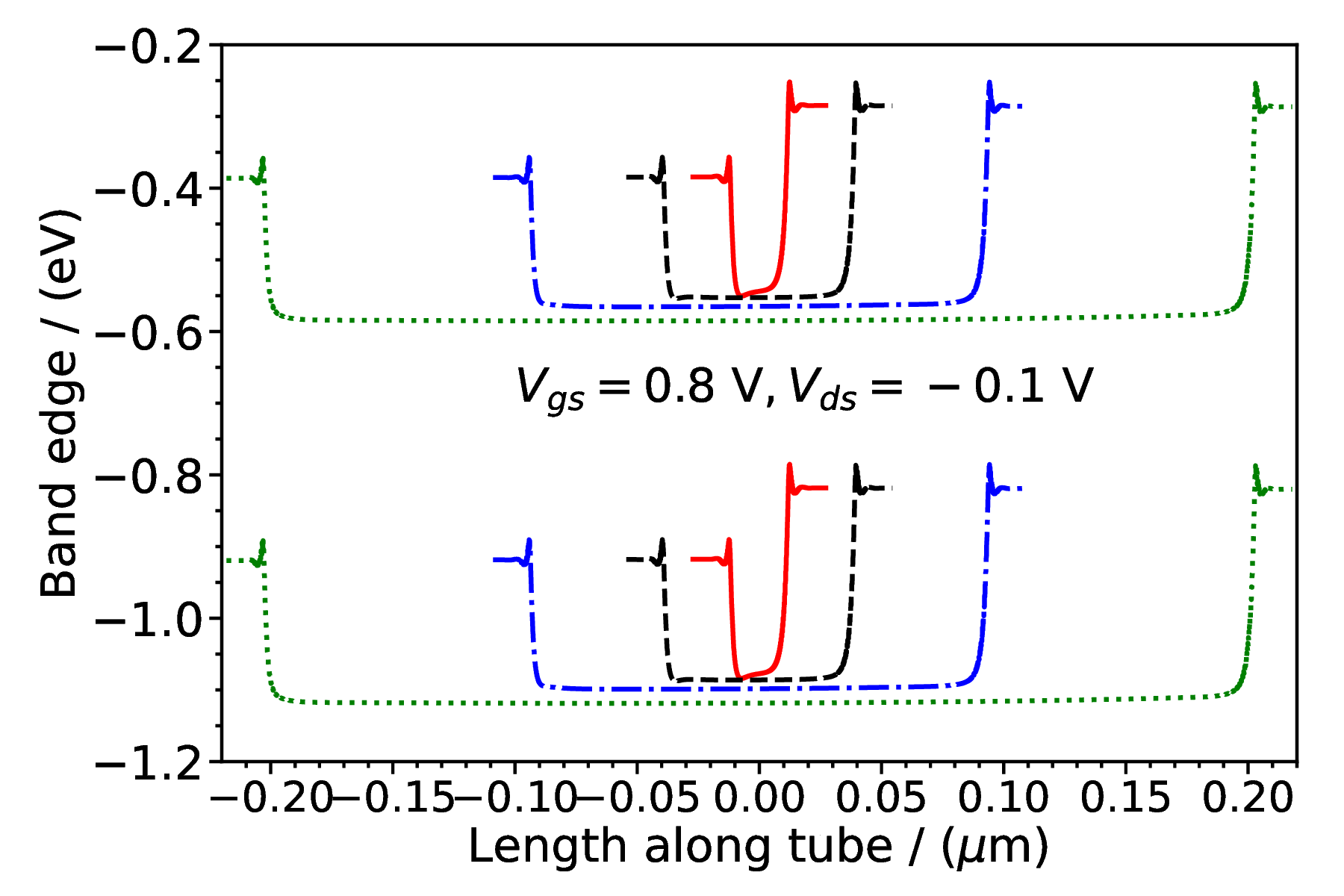}}}\,
        \sidesubfloat[]{\label{f:integrand}{\includegraphics[width=0.45\textwidth]{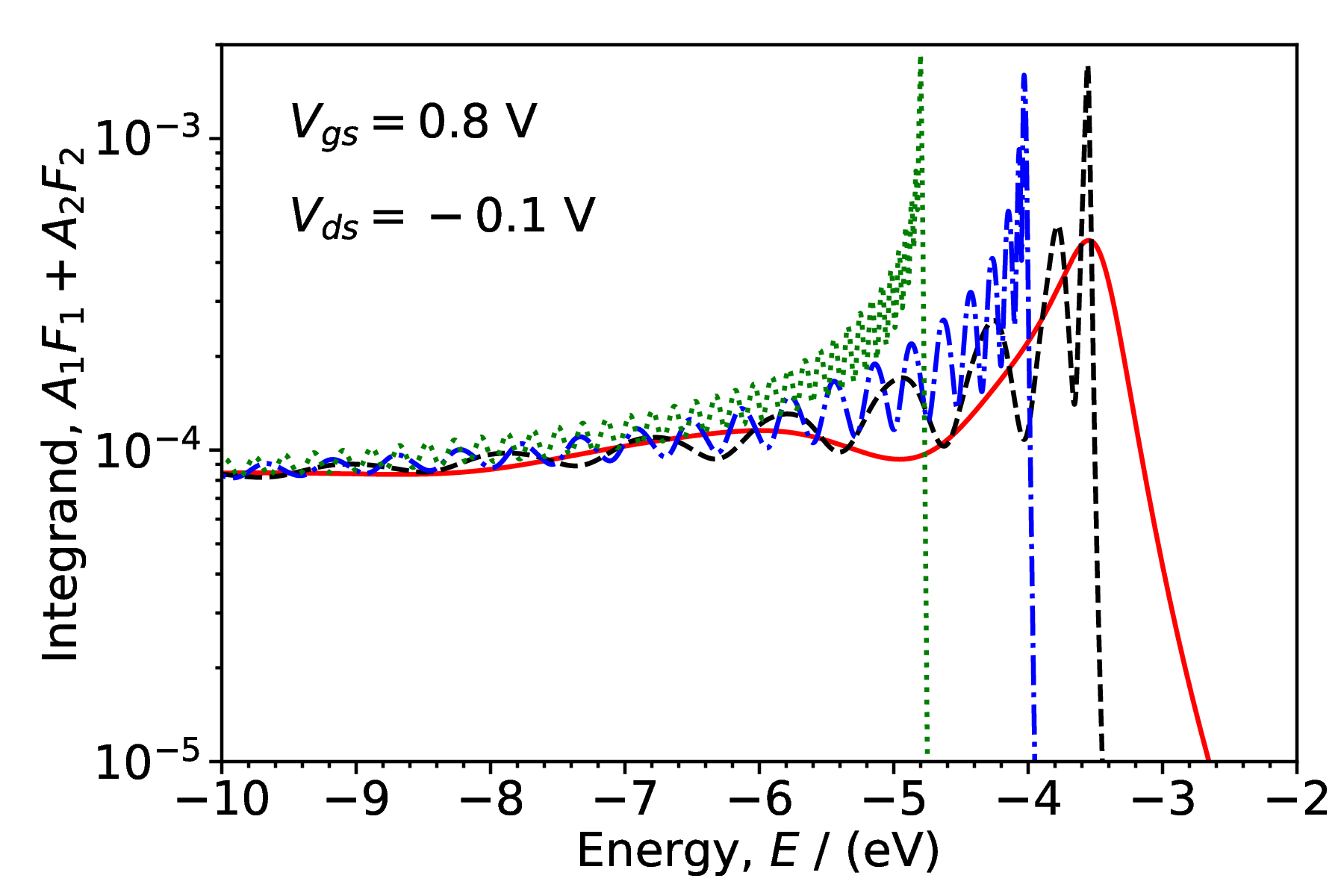}}}\,
        \caption{(\textit{a}, \textit{b}) Current-voltage characteristics for four channel lengths analyzed for parallel performance studies. (\textit{c}) Comparison of band bending across the four cases.  (\textit{d}) Integrand at the channel center for nonequilibrium charge density calculation. Legends for all figures are shown in (\textit{a}), representing channel lengths.}
    \label{f:Scaling}
\end{figure}

Figures~\ref{f:WeakResults_I_Vds} and \ref{f:WeakResults_I_Vgs} depict the current-voltage characteristics for the four simulated cases, while Figure~\ref{f:WeakResults_bandStructure} illustrates the band bending at $V_{gs}=0.8~$V and $V_{ds}=-0.1~$V. 
Initially, we compute the $I-V_{ds}$ characteristics at $V_{gs}=0.8~$V when the transistor is nearly in the OFF state.
Next, utilizing the charge density profile at $V_{ds}=-0.1~$V, we begin simulations to compute the $I-V_{gs}$ characteristics, sweeping $V_{gs}$ from $0.8~$V to $-0.3~$V while using the converged solution at the previous $V_{gs}$ to initialize the computation at the new $V_{gs}$. 

In the subthreshold region the $I-V_{ds}$ characteristics reveal a p-type FET behavior with saturation of the current at ~-0.1 V. The increase in the drain-source saturation current magnitude with decreasing channel length is a consequence of the DIBL discussed earlier. Of note is that the impact of DIBL on the source-drain current is observed even when the channel is much longer than the oxide thickness of 10 nm. This impact of the channel length is also obversed in the subthreshold region of the $I-V_{gs}$ characteristics, while the ON state current remains comparable to that of the shortest channel length.

Turning to the question of integration points, we note that the band structure of an infinitely long carbon nanotube exhibits  van hove singularities at the band edges.
For a finite CNT, as the length gets longer, the singularity gets sharper, requiring more integration points for accurate calculation of the second part of the integral to compute charge density, shown in equation~(\ref{eq:Rho}).
To illustrate, the integrand $A_1 F_1 + A_2 F_2$ at the center of the channel is shown in Figure~\ref{f:integrand} as a function of energy, $E$.
We observe that as the channel length increases, not only does the singularity get sharper, but it also exhibits oscillating structures on the lower energy side.

Therefore, to reduce the number of integration points, we use a semi-adaptive method and breakdown the integration path from $E_{min}$ to $E_{max}$ into three subparts, $E_{min}$ to $E_{l1}$, $E_{l1}$ to $E_{l2}$, and $E_{l2}$ to $E_{max}$.
In each of these subregions we can specify the density of integration points, defined by the number of integration points per $kT$.
Before starting the simulation, we do not know the exact location of the singularity, and therefore, we start the simulation with a guess for $E_{l1}$ and $E_{l2}$, compute the integrand at the center of the channel as a function of energy, obtain the peak of the singularity $E_{s}$, and refine limits as $E_{l1} = E_{l1} - akT$, $E_{l2} = E_s + bkT$, where constants $a$ and $b$ can be specified by the user, typically set to $3$ and $1$, respectively.
The number of integration points change according to the density of integration points set by the user in each of these subregions.
%The charge is computed with the updated limits.
Note that during the self-consistency iterations the location and peak of singularity may change, and therefore, we update these limits at the first iteration for a given set of conditions and update it at a set period of 50 iterations.
For the evaluation of integration in each subregion, we use Gauss-Legendre quadrature to further improve the accuracy. 
For instance, to simulate the largest channel length we used densities of integration points of (500, 1600, 50), respectively, which lead to approximately 12,000 integration points in total. For the performance analyses described above, we kept this total number of integration points constant for consistent scaling of the workload across various cases. However, in reality, to obtain a numerically accurate solution at 23.9, 78.4, and 187.4 channel lengths we only need approximately 400, 1000, and 6000 total integration points, respectively. In fact, the smallest channel length can be simulated with just one integration path from $E_{min}$ to $E_{max}$.
We suspect that a fully adaptive scheme might be necessary for simulating micron-scale CNTs self-consistently.

%\subsection{Application to 3D CNTFET devices}
%\label{sec:multipleCNTs}

%\tr{This section demonstrates the application of ELEQTRONeX to model computationally challenging configurations of CNTFETs. In \S \ref{sec:scaling_results} we discuss the results obtained from the nonequilibrium simulations run for the parallel performance analyses with gate-all-around configurations, discussed in \S \ref{sec:scaling} in Table \ref{tab:weak-scaling}, as well as some considerations in choosing integration paths and points to obtain converged self-consistent solution for longer nanotubes.} In \S \ref{sec:nonpllPlanar}, we study the effect of modeling multiple CNTs in a fully 3D planar CNTFET configuration and study the impact of misalignment of CNTs compared to perfectly parallel alignment simulated with and without periodic boundary conditions.

\subsection{3D Nonparallel Planar CNTFET Configurations}
\label{sec:nonpllPlanar}

In \S \ref{sec:validation}, Figure~\ref{f:L10nm_periodic_G} showed the conductance from a planar CNTFET configuration with periodic boundary conditions in the lateral direction, effectively emulating an infinite array of CNTs spaced apart by 3.2 nm. Although this setup is instructive for theoretical analysis, experimental devices typically contain a finite number of CNTs in the channel, necessitating 3D non-periodic simulations to reflect more realistic conditions. To showcase the capabilities of our approach, we simulated five nanotubes in the channel by applying Neumann boundary conditions instead of periodic ones. An additional spacing of $d$ was introduced between the lateral boundaries and the outermost nanotubes.

\begin{figure}[H]
    \centering
        \sidesubfloat[]{\label{f:G_1x_5short}{\includegraphics[width=0.45\textwidth]{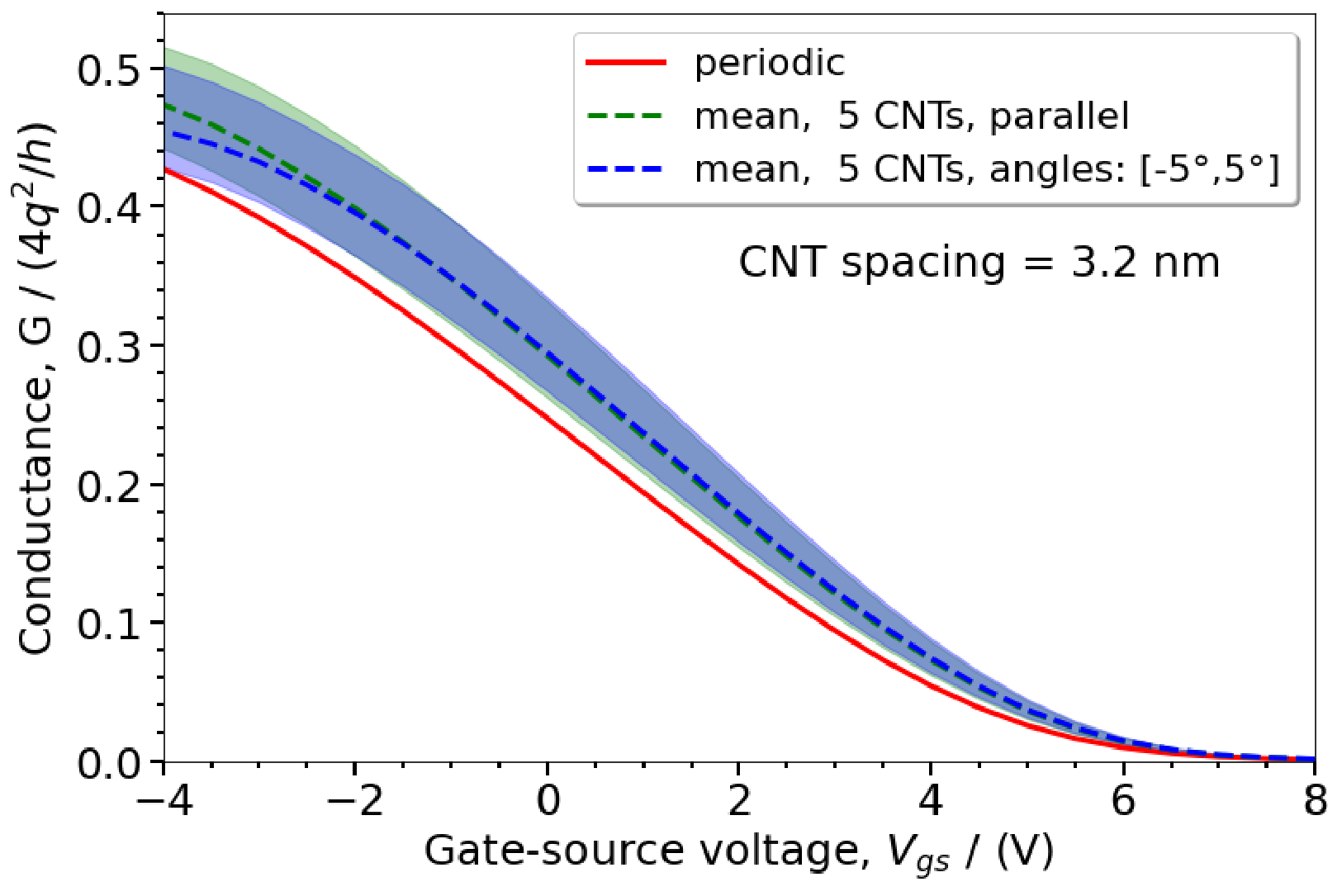}}}
        \sidesubfloat[]{\label{f:G_4x_20short}{\includegraphics[width=0.45\textwidth]{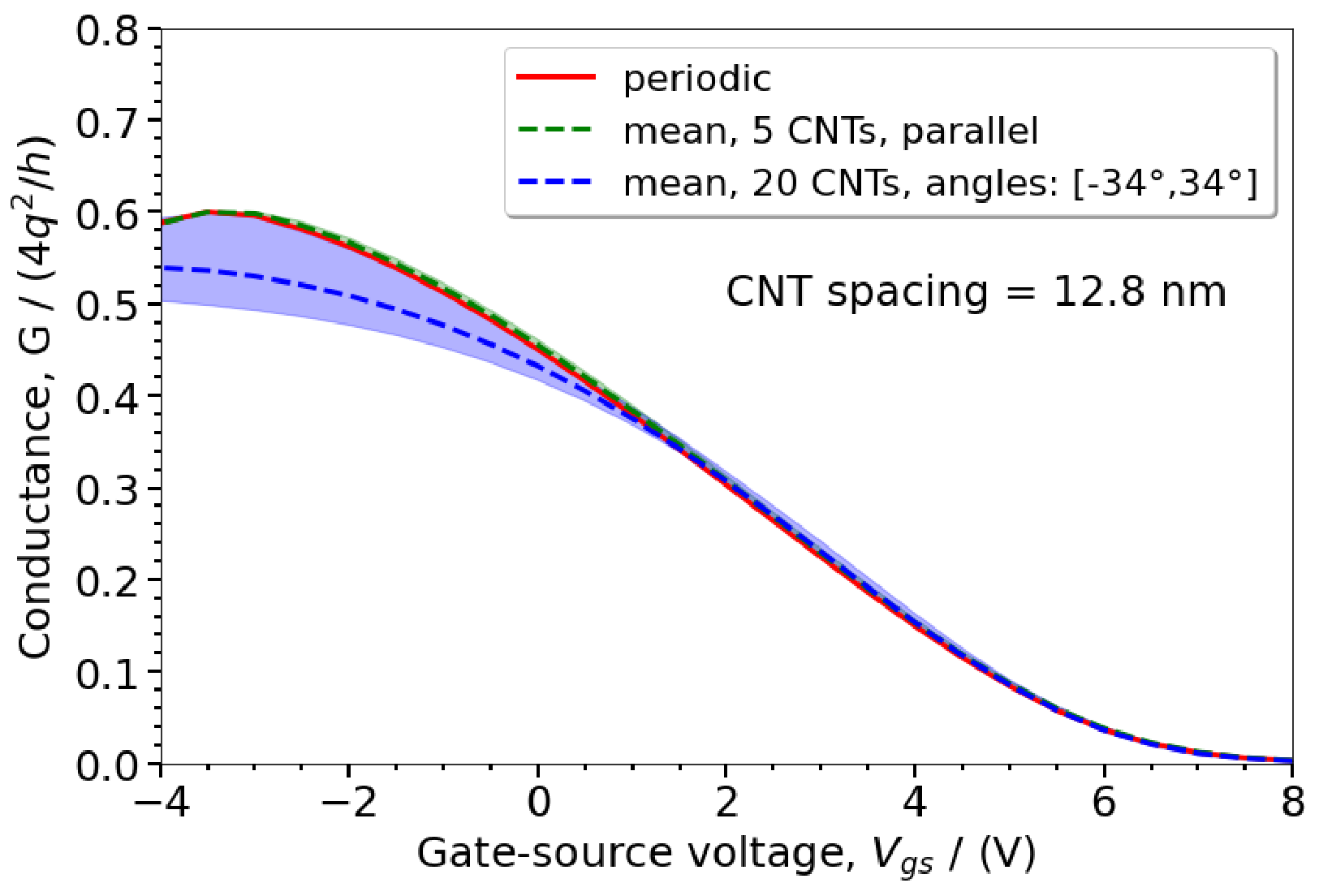}}}\,
        \sidesubfloat[]{\label{f:G_1x_5short_phi}{\includegraphics[width=0.45\textwidth]{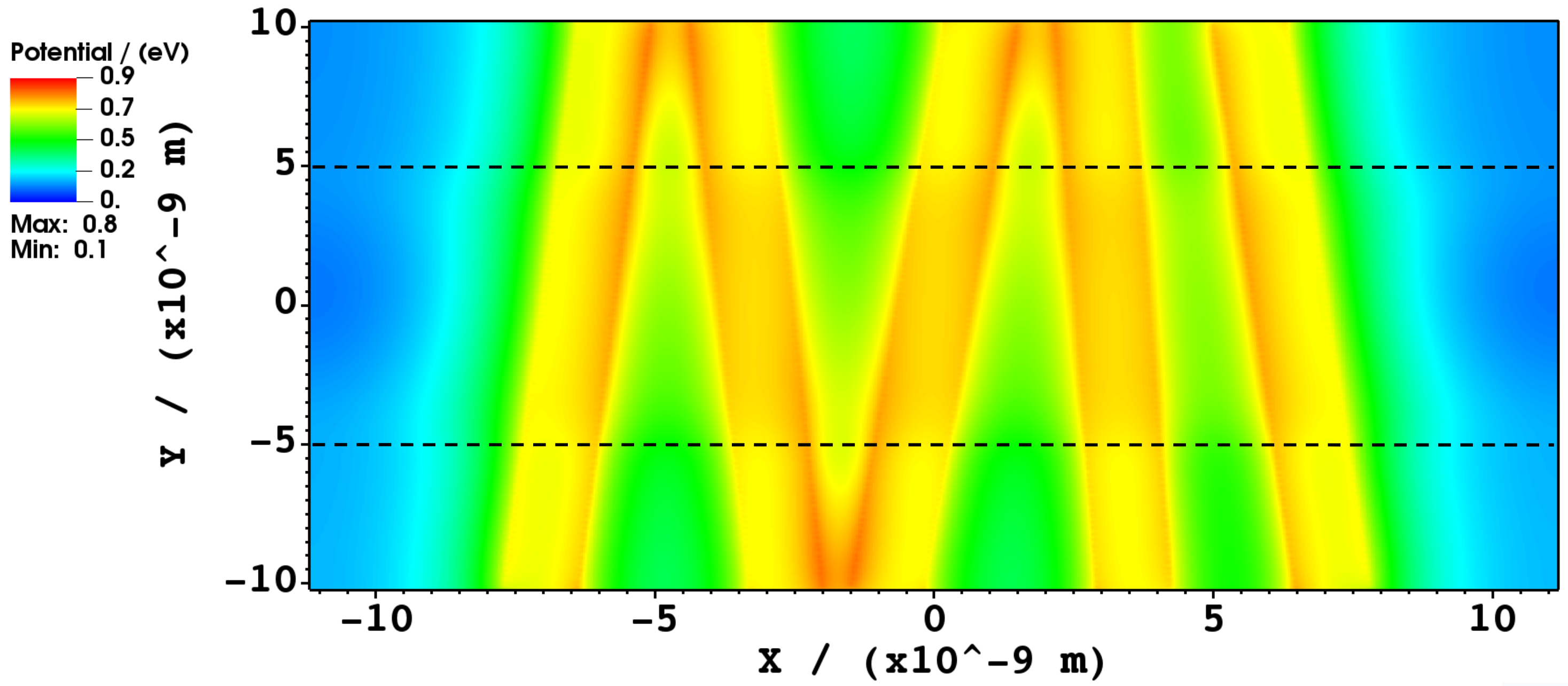}}}
        \sidesubfloat[]{\label{f:G_4x_20short_phi}{\includegraphics[width=0.45\textwidth]{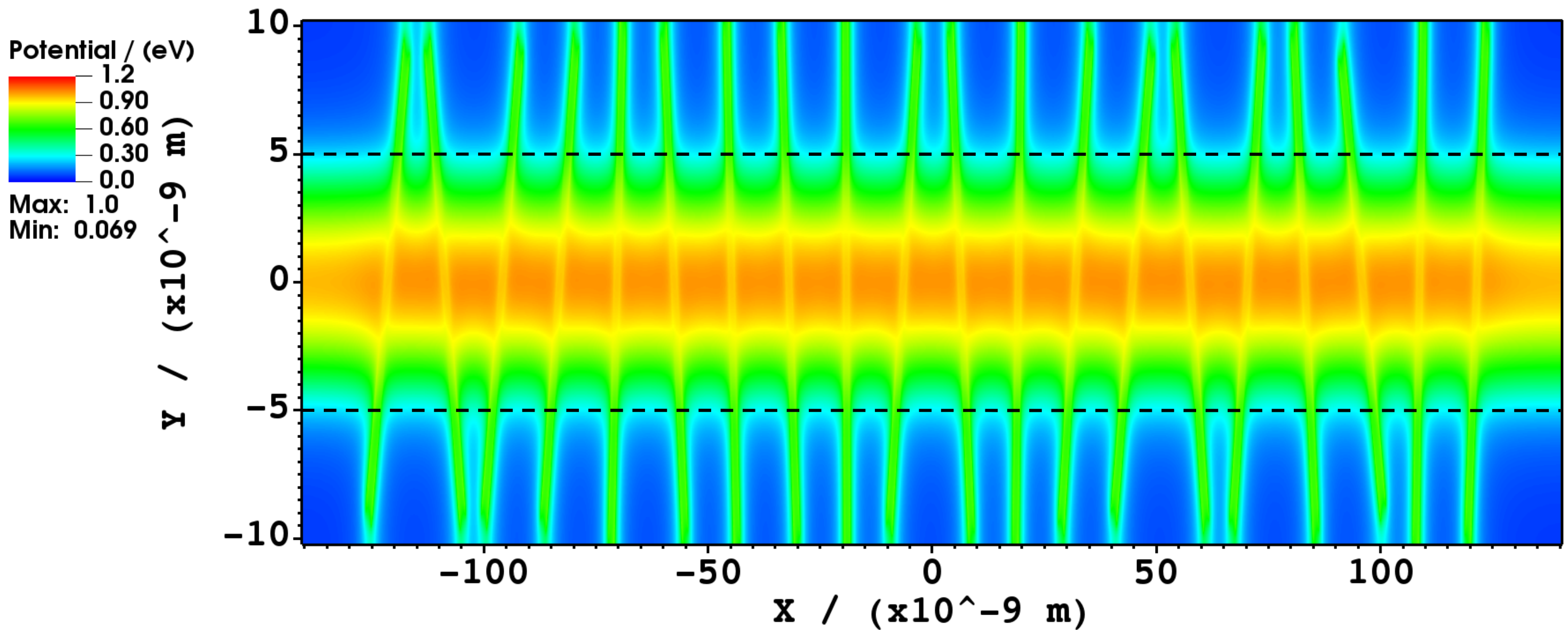}}}\,        
    \caption{Comparison of zero-bias conductance $G$ versus gate-source voltage $V_{gs}$ for periodic, non-periodic and parallel, and non-parallel configurations with different CNT spacing $d$, and channel length $L$. (\textit{a}) $d=3.2$~nm, $L=10$~nm, (\textit{b}) $d=12.8$~nm, $L=10$~nm. For nonperiodic simulations, we show the mean conductance with the colored spread around it marking the maximum and minimum conductance for that configuration. (\textit{c}) and (\textit{d}) show electrostatic potential in non-parallel configurations at $V_{gs}=-2$~V.}
\end{figure}

The simulation results, compared in Figure~\ref{f:L10nm_periodic_G}, show that the conductance profiles of the middle three nanotubes closely resemble those obtained with periodic boundary conditions. However, the conductance of the outermost nanotubes, labeled as `second adjacent,' is significantly higher. These results are shown again in Figure~\ref{f:G_1x_5short}, where we present the conductance obtained from the non-periodic configuration as a shaded region around the mean conductance to provide a clearer visual representation of variability. The finite CNT case carries more current at a given gate voltage,  $11.5\%$ more current in the ON state (2.9 mS/$\mu$m vs 2.6 mS/$\mu$m), and a reduced substhrehsold swing (from 2986 mV/dec to 2011 mV/dec)  due to the different gate capacitive coupling.

Next, in the same Figure~\ref{f:G_1x_5short}, we also compare the impact of CNT misalignment--a common occurrence in CNT-array devices as noted in various studies \cite{Arnold,Peng}.
In this configuration, 5 CNTs are rotated around their midpoint by angles sampled randomly in the range [$-5^{\circ}$, $5^{\circ}$], such that CNTs do not touch or overlap other for this 10 nm channel case, as seen from the contours of electrostatic potential in Figure~\ref{f:G_1x_5short_phi}.
Such configurations lack periodicity, requiring fully 3D simulations.

It is seen that the conductance for the non-parallel configuration only differs slightly from the parallel configuration  when the FET is in the ON state (2.8 mS/$\mu$m vs 2.9 mS/$\mu$m). This small difference is attributed to the small range of possible angles due to the short channel and close proximity between CNTs. 
In addition, the CNT misalignment has little impact on subthreshold swing (2010 mV/dec vs 2011 mV/dec) or threshold voltage compared to the parallel 5 CNT case, while the variation in the variation of the subthreshold swing from each CNT is only about 70 mV/decs.

\begin{figure}[H]
    \centering
        \sidesubfloat[]{\label{f:G_4x_20long}{\includegraphics[width=0.45\textwidth]{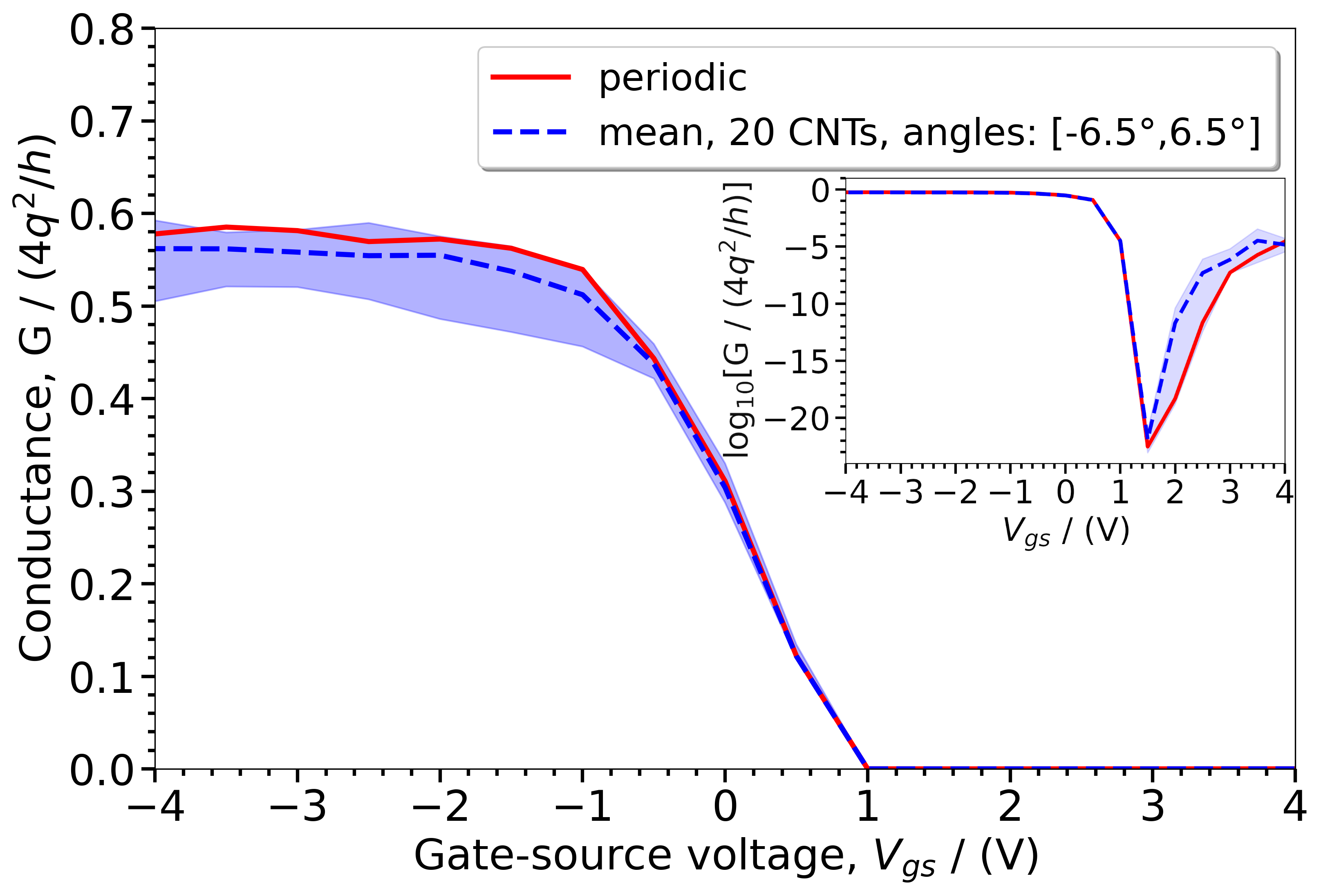}}}
        \sidesubfloat[]{\label{f:G_4x_20long_spacing}{\includegraphics[width=0.45\textwidth]{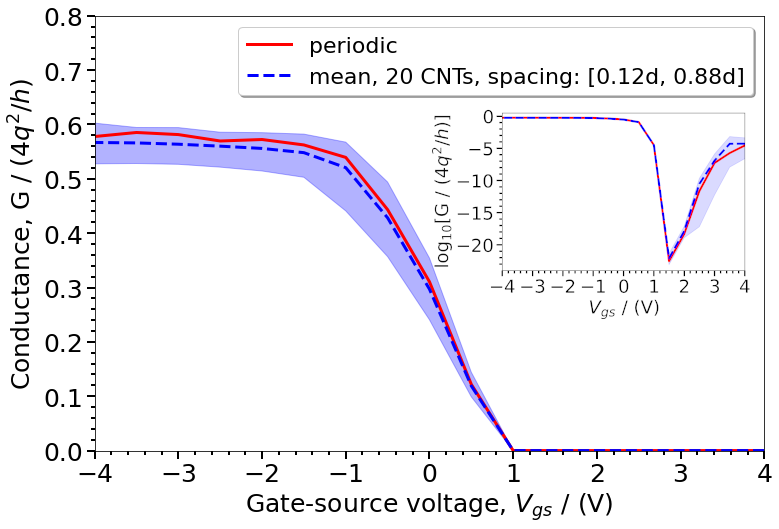}}}\,
        \sidesubfloat[]{\label{f:phi_20long_4x}{\includegraphics[width=0.44\textwidth]{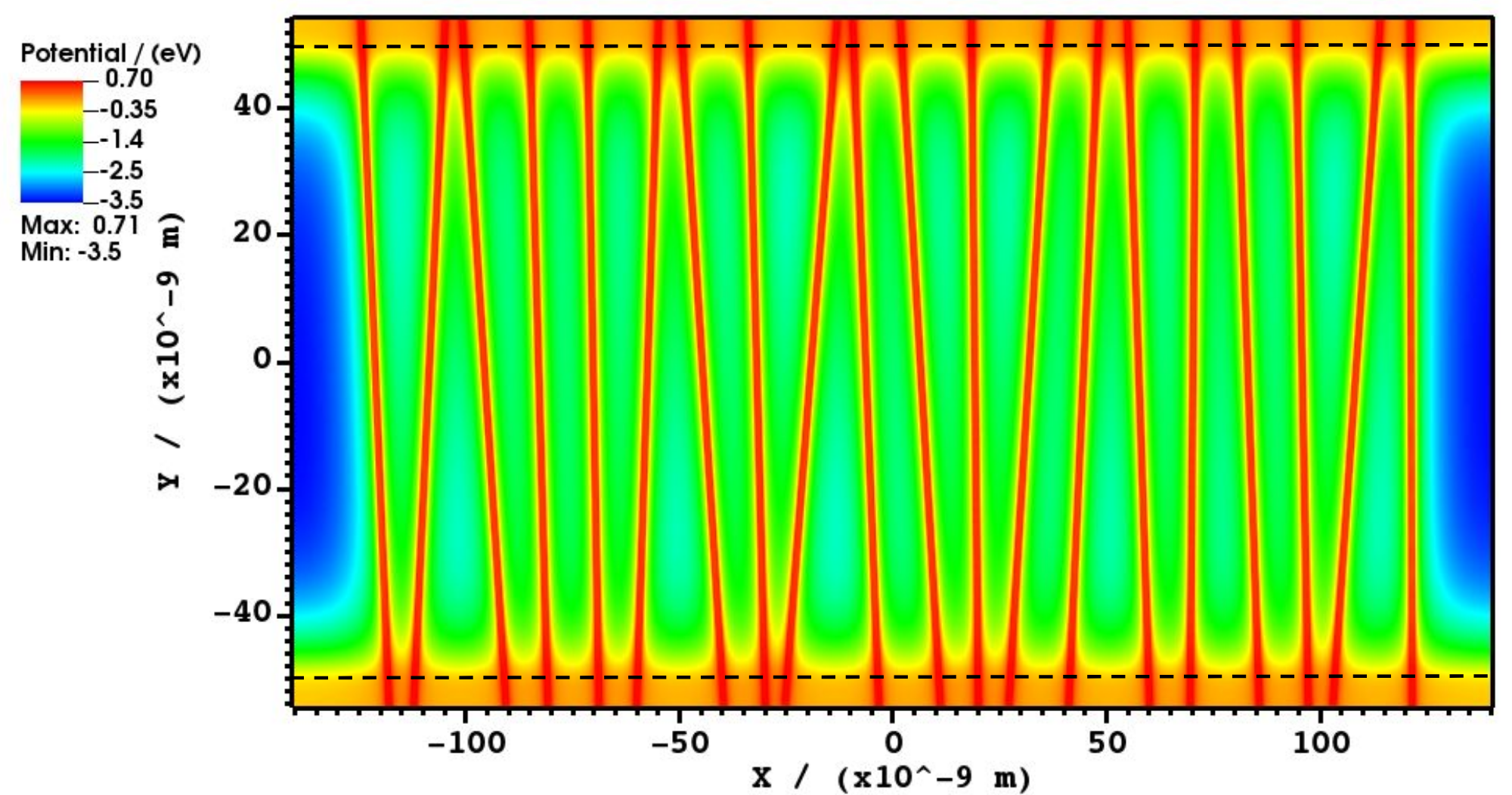}}}\hfill
        \sidesubfloat[]{\label{f:phi_20long_4x_spacing}{\includegraphics[width=0.44\textwidth]{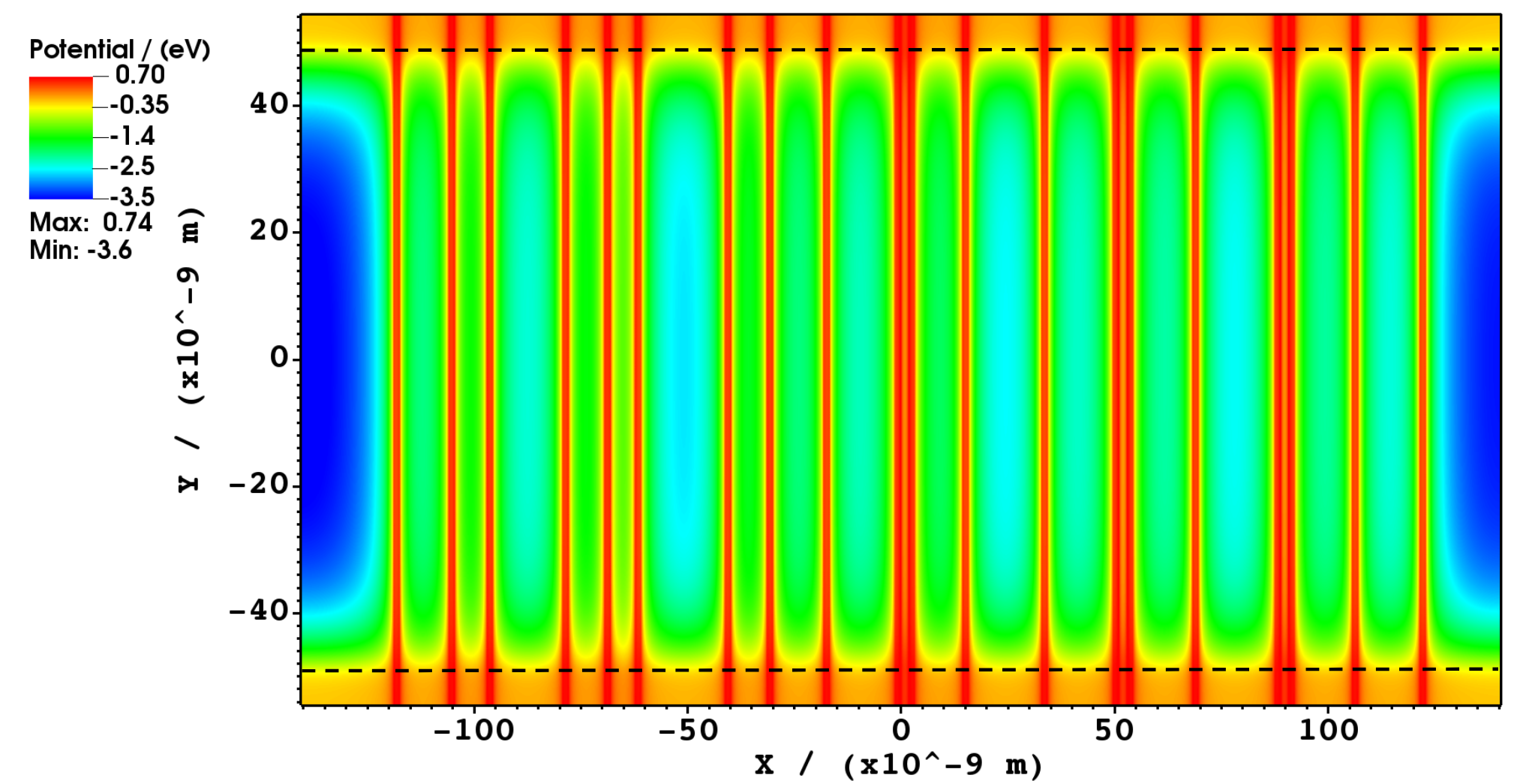}}}\,
        \sidesubfloat[]{\label{f:EvEc_20long_4x}{\includegraphics[width=0.44\textwidth]{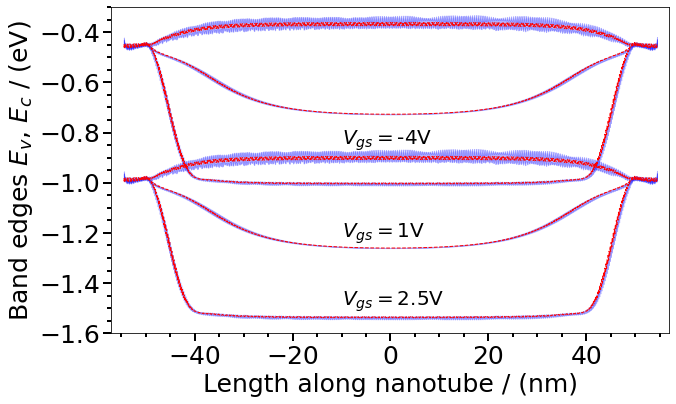}}}\hfill
        \sidesubfloat[]{\label{f:EvEc_20long_4x_spacing}{\includegraphics[width=0.44\textwidth]{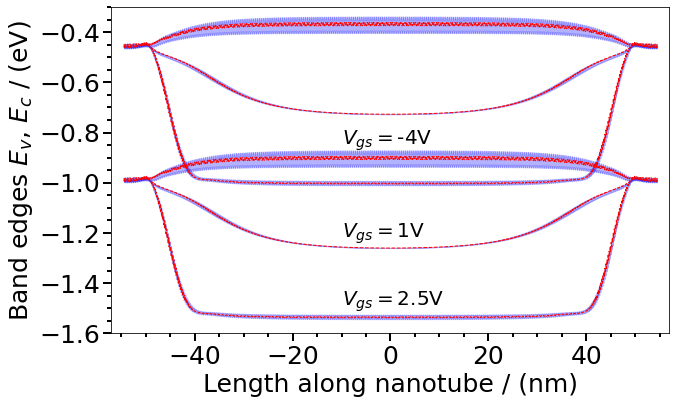}}}\,
    \caption{Planar configurations of 20 CNTs with an average spacing of $d=12.8$~nm and a channel length of $L=100~$nm, illustrating the effects of variations in non-parallel orientation angles and non-uniform spacing on key characteristics compared to a periodic arrangement. For angular variations: (\textit{a}) comparison of $G$ versus $V_{gs}$, with inset showing the same plot on a $y-log$ scale, (\textit{c}) electrostatic potential at $V_{gs}=-4$~V on a plane passing through CNTs, (\textit{e}) comparison of band structure. For spacing variations: similar characteristics are shown in (\textit{b}), (\textit{d}), and (\textit{f}), respectively.}
\end{figure}
 
Next, in Figure~\ref{f:G_4x_20short}, we study the effect of modeling a finite number of parallel and non-parallel nanotubes when nanotubes are spaced farther apart ($d$=12.8~nm) such that $d$ is larger than the gate-oxide thickness of 10~nm.
In this case, contrary to the simulation with the smaller CNT spacing, the finite array of 5 parallel CNTs does not show much deviation from the corresponding infinite array because the CNT-CNT cross-talk is screened by the gate. (ON stage conductance 3.6 mS/$\mu$m in both cases, subthreshold swing 1778 mV/dec vs 1785 mV/dec).
However, for the case with misalignment with angles in the range [$-34^{\circ}$, $34^{\circ}$], we observe a decrease of the average ON state conductance by about 8.3\% to 3.4 mS/$\mu$m
This decrease is caused by cross-talk between the CNTs when the spacing between them decreases for part of the channel length, as seen from the contours of electrostatic potential in Figure~\ref{f:G_4x_20short_phi}.

To understand how the amount of variation in conductance changes with channel length, we increased the channel length to 100~nm and modeled 20 CNTs with an average spacing of $d$=12.8~nm and angular rotation within the range [$-6.5^{\circ}$, $6.5^{\circ}$].
Figures~\ref{f:G_4x_20long}, \ref{f:phi_20long_4x}, and \ref{f:EvEc_20long_4x} show conductance, electrostatic potential, and band structure for this simulation.
From the conductance plot, we again note that the misalignment results in only about 3\% lower conductance with respect to a periodic arrangement when the channel is ON (3.4 mS/$\mu$m vs 3.5 mS/$\mu$m), and has no impact on the subthreshold swing (76.7 mV/dec in both cases).

The insert shows the conductance on a y-log scale, revealing a minimum in the conductance and an increase in the current for higher gate-source voltages. 
This increase is caused by band-to-band tunneling as a result of the conduction band-edge approaching the Fermi level at -1 eV.
We further note that the band-structure obtained from non-parallel simulation differs from a periodic simulation only when the channel is ON; when there is more charge in the channel, we see more cross-talk resulting in slightly lower conductance.

The above results suggest that CNT array devices have good resiliency to misalignment over a broad parameter range. This can be understood primarily based on screening arguments: CNT-CNT interactions due to misalignment occur primarily near the source and drain electrodes, but are effectively screened by the contact metal and the gate. 

We also explored the effect of non-uniform spacing between nanotubes while keeping them parallel.
The average spacing is kept constant at $d=12.8$~nm, and the spacing between each nanotube is varied in the range [0.12d, 0.88d].
Figures~\ref{f:G_4x_20long_spacing}, \ref{f:phi_20long_4x_spacing}, and \ref{f:EvEc_20long_4x_spacing} show conductance, electrostatic potential, and band structure for this case.
This simulation showed a broader variation in the conductance of individual CNTs compared to a periodic simulation, starting from the linearly increasing region of conductance ($V_{gs}=1~V$ to $-1~V$).
Yet, the total conductance of the device is close to that of the periodic system; the maximum variation seen is still only about 3\% when the channel is ON.
The band-structure shows a maximum variation with respect to periodic configuration, when the channel is ON. Variations in the subthreshold swing or the threshold voltage are also minimal in this case.

Another factor of importance for digital electronics is the CNTFET OFF current, which has been shown to be dominated by phonon-assisted band-to-band tunneling \cite{Wong}. While we plan to add electron-phonon scattering in the future, here we provide a simple estimate based on the direct tunneling current. In CNTs, the photon-assisted tunneling current is dominated by scattering with optical phonons of energy 0.2 eV \cite{Koswata}. Thus, we found the gate-source voltage at which the conduction band edge is 0.2 eV above the lead Fermi level and estimated the OFF state conductance from the conductance calculated at this value without electron-phonon scattering. The results can be found in Table~\ref{tab:summary}. While the values are approximate, they show that the OFF state current is not strongly affected by the finite size of the array and misorientation.

%The results of our simulations can be found in Table~\ref{tab:weak-scaling-equilibrium}. We assessed the OFF current by taking the direct band-to-band tunneling current at V$_{gs}$ = 2V. In principle, phonon-assisted tunneling would lead to an increase of the current before this value, but to first order this is proportional to the direct tunneling current; our assessment is meant to convey trends in the impact of non-idealities rather than quantitative values.

%The results of our simulations can be found in Table~\ref{tab:weak-scaling-equilibrium}. We calculated the subthreshold swing such at $V_{gs}$ values, when the barrier between the Fermi level and valence band-edge in the channel goes from 5 to 8~kT. \tr{We assessed the OFF current when the conduction band edge in the channel is 0.2 eV above the Fermi level (-1 eV) to incorporate the effect of phonon-assisted tunneling. Although, we have not modeled electron-phonon scattering in this work,} our assessment is meant to convey trends in the impact of non-idealities rather than quantitative values. 

\begin{table}[H]
\begin{threeparttable}
\caption{Impact of non-idealities on planar CNTFETs$^*$.}
\label{tab:summary}       % Give a unique labela
        \begin{tabular}{|l|l|l|l|l|l|c|}
\hline\noalign{\smallskip}
L    & d    & Geometry               & $G_{ON}$ & SS  & $G_{OFF}$ \\
(nm) & (nm) &                        & ($m$S/$\mu$m) & (mV/dec) & ($\mu$S/$\mu$m)\\
\hline\noalign{\smallskip}
10   & 3.2  & parallel, infinite     & 2.6               & 2986   & 6.7 \\
10   & 3.2  & parallel, 5 CNTs       & 2.9 $\pm$ 0.21   & 2011 $\pm$ 66  & 6.9 $\pm$ 0.4 \\
10   & 3.2  & misaligned, 5 CNTs     & 2.8 $\pm$ 0.17   & 2010 $\pm$ 70  & 6.8 $\pm$ 0.4  \\
\hline\noalign{\smallskip}
10   & 12.8 & parallel, infinite     & 3.6               & 1785   & 1.8 \\
10   & 12.8 & parallel, 5 CNTs       & 3.6 $\pm$ 3.2e-3  & 1778 $\pm$11 & 1.8 $\pm$ 3.5e-3 \\
10   & 12.8 & misaligned, 20 CNTs     & 3.3 $\pm$ 0.2    & 1668 $\pm$56 & 1.4 $\pm$ 0.34 \\
\hline\noalign{\smallskip}
100  & 12.8 & parallel, infinite     & 3.5               & 76.7  & 3.9e-4 \\
100  & 12.8 & misaligned, 20 CNTs    & 3.4 $\pm$ 0.11   & 76.7 $\pm$ 2.0  & 3.7e-4 $\pm$ 4.6e-5\\
100  & 12.8 & varying pitch, 20 CNTs & 3.4 $\pm$ 0.15   & 77.2 $\pm$ 1.0  & 4.2e-4 $\pm$ 4.4e-5 \\
\noalign{\smallskip}\hline\noalign{\smallskip}
\end{tabular}
    \begin{tablenotes}
    \item [$^{*}$] Oxide thickness was kept constant to 10 nm. For cases with multiple nanotubes, one standard deviation variation is shown.
    \end{tablenotes}
\end{threeparttable}
\end{table}

In terms of computational cost, the simulations with 20 CNTs employed 10560, 768, and 768 cells in the X, Y, and Z directions, respectively, with a cell size of 0.026625 nm in each direction.
The computation utilized 512 MPI ranks each associated with one GPU, averaging approximately 3.6~s per iteration for electrostatics and 0.65 seconds per iteration with 90 integration points for NEGF, while the time required for other steps in the iteration was an order of magnitude lower.
The code required 42 iterations on average, i.e. approximately 3 minutes per gate-source voltage condition.

% conclusion
%\clearpage
\section{Conclusions}
\label{sec:conclusion}

In summary, we present a robust, portable, and open-source implementation of the NEGF method self-consistently coupled with electrostatics, employing the AMReX library and MPI/GPU parallelization strategies to leverage modern supercomputing architectures.
We demonstrate rigorous validation of our code against established literature for computing DC transport properties across various CNTFET configurations, including multi-nanotube simulations.

The performance studies underscore ELEQTRONeX's computational efficiency, particularly in modeling larger systems such as long device channels and larger computational domains for electrostatics.
The time complexity of the multigrid electrostatics is $O(N)$ for a system of size $N$, while the parallel implementation achieves a near-ideal $O(1)$ time complexity, bounded only by inter-processor communication.
The time complexity of the serial NEGF algorithm is $O(N^2)$, yet we demonstrate that it can achieve $O(N)$ performance with heterogeneous MPI and GPU implementation, while optimizations such as asynchronous CPU-to-GPU copying can further halve the scaling factor.
Additionally, we introduce an MPI/GPU parallel implementation of Broyden's modified second algorithm for self-consistency, enhancing parallel time complexity to $O(m_{\rm avg})$ compared to $(Nm_{\rm avg})$ serial implementation, where $m_{\rm avg}$ represents average number iterations required for convergence.

Furthermore, we demonstrate our capability to model fully 3D non-periodic configurations of CNTFETs and investigate the accuracy of simple periodic calculations compared to realistic experimental setups.
Such setups may involve a finite number of nanotubes that are not perfectly parallel or equally spaced. Across a broad range of configurations and device geometries we find that CNTFETs are robust against misalignment and varying pitch for several key device performance metrics.

In the future we plan to add several additional capabilities to ELEQTRONeX including electron-phonon interactions, trap charges, top-gate geometries, SRAM cell configurations, bulk materials, and time-dependent NEGF to address computationally challenging state-of-the-art problems.

\section*{Acknowledgements}
Work by S. S.~Sawant and Z.~Yao was supported by the U.S.~Department of Energy, Office of Science, Office of Applied Scientific Computing Research, the Microelectronics Co-Design Research Program, under Contract DE-AC02-05-CH11231 (Co-Design and Integration of Nano-sensors on CMOS).
Work by F.~L\'eonard was supported by the U.S.~Department of Energy, Office of Science, the Microelectronics Co-Design Research Program, under Contract DE-NA-0003525 (Co-Design and Integration of Nano-sensors on CMOS). Sandia National Laboratories is a multimission laboratory managed and operated by National Technology and Engineering Solutions of Sandia, LLC., a wholly owned subsidiary of Honeywell International, Inc., for the U.S.
Department of Energy’s National Nuclear Security Administration under contract DE-NA-0003525.
Work by A.~Nonaka was supported by the Exascale Computing Project (17-SC-20-SC), a collaborative effort of the U.S. Department of Energy Office of Science and the National Nuclear Security Administration; the U.S. Department of Energy, Office of Science, Office of Advanced Scientific Computing Research, Exascale Computing Project under contract DE-AC02-05CH11231.
This research used resources of the National Energy Research Scientific Computing Center (NERSC), a Department of Energy Office of Science User Facility using NERSC award ASCR-ERCAP0026882.

Authors would also like to thank Dr.~Dmitri Nikonov for valuable discussions during the development of the solver.
\bibliographystyle{elsarticle-num} 
\bibliography{references}

\end{document}